\documentclass[12pt,letterpaper]{article}

\usepackage{amsfonts}
\usepackage{amssymb}
\usepackage{graphicx}
\usepackage{latexsym}
\usepackage{epstopdf}

\def\ee{\end{eqnarray}}

\def\d{\partial}
\def\nn{\nonumber}

\newcommand{\half}{{\textstyle{\frac12}}}

\newcommand{\be}{\begin{eqnarray}}
\newcommand{\en}{\end{eqnarray}}
\newcommand{\bea}[1]{\left(\begin{array}{#1}}
\newcommand{\ena}{\end{array}\right)}
\newcommand{\CO}{\mathcal{O}}
\newcommand{\CH}{\mathcal{H}}

\begin{document}

\vspace{5mm}
\vspace{0.5cm}
\begin{center}

\def\thefootnote{\fnsymbol{footnote}}

{\Large \bf Contributions to the Dark Matter 3-Point Function
from the Radiation Era}
\\[0.5cm]
{\large  A.~Liam Fitzpatrick$^{1,2}$, Leonardo Senatore$^{2,3,4}$, and 
Matias Zaldarriaga$^{2,3}$}

{\small \textit{$^{1}$
Physics Department \\
Boston University, Boston, MA 02215, USA}}

{\small \textit{$^{2}$
Jefferson Physical Laboratory \\
Harvard University, Cambridge, MA 02138, USA}}

{\small \textit{$^{3}$
Center for Astrophysics \\
Harvard University, Cambridge, MA 02138, USA}}

{\small \textit{$^{4}$ 
School of Natural Sciences \\
 Institute for Advanced Study,
Olden Lane, Princeton, NJ 08540, USA}}

\end{center}

\vspace{.8cm}

\hrule \vspace{0.3cm}
{\small  \noindent \textbf{Abstract} \\[0.3cm]
\noindent
We consider the contribution to the three-point function of
matter density fluctuations from nonlinear growth after modes
re-enter the horizon, and discuss effects that must be included
in order to predict the three-point function with an accuracy
comparable to primordial nongaussianities with $f_{\rm NL} \sim
few$.  In particular, we note that the shortest wavelength modes 
measured in galaxy surveys entered the horizon during the radiation
era, and, as a result, the radiation era modifies their three-point
function by a magnitude equivalent to $f_{\rm NL} \sim \CO(4)$.
On longer wavelengths, where the radiation era is negligible,
we find that the corrections to the nonlinear growth from 
relativistic effects become important at the level $f_{\rm NL} \sim
few$.  We implement a simple
method for numerically calculating the three-point function,
by solving the second-order equations of motion for the perturbations
with the first order perturbations providing a source.  

\vspace{0.5cm}  \hrule
\def\thefootnote{\arabic{footnote}}}
\setcounter{footnote}{0}

\section{Introduction}

The spectrum of initial density perturbations from inflation is known
to be nearly gaussian.  Should deviations from a gaussian distribution be 
detected, however, they will imply potentially powerful constraints
on models of inflation.  In particular, the shape and size  of the
three-point function of primordial density perturbations 
in many models is predicted to be near current experimental limits, and is one
of the best observables for distinguishing different inflationary models
\cite{Maldacena:2002vr,Acquaviva:2002ud,Arkani-Hamed:2003uz,Alishahiha:2004eh,Chen:2006nt,Zaldarriaga:2003my,Creminelli:2006gc,us,Creminelli:2006xe,Armendariz-Picon:1999rj,Stewart:1993bc,Arkani-Hamed:2003uy} as well as its
(perhaps less compelling) alternatives \cite{Creminelli:2007aq}. 
The best measurements so far come from cosmic microwave background (CMB) data
 \cite{Spergel:2006hy,Creminelli:2006rz,Babich:2004yc}, but large-scale structure
(LSS) measurements are improving and will become comparable. 
The shortest wavelength measured by LSS is smaller than for the CMB,
and it is a three-dimensional map instead of a two-dimensional one.
Therefore LSS measurements potentially 
include many more modes.
However, they also therefore involve modes that
entered the horizon at earlier times, when the universe was radiation dominated.
Our purpose is to quantify the effect of radiation on nongaussianities
in LSS measurements in comparison to that from
primordial nongaussianities.

The significance of the radiation era is due to the nonlinear growth of
nongaussianities, which occurs even for gaussian initial
conditions \cite{Sefusatti:2007ih,Bernardeau:2001qr}.  The leading contributions have been studied in the 
perturbation theory (PT) formalism, dropping the radiation component
of the universe and using Newtonian gravity \cite{Goroff:1986ep,Jain:1993jh,Makino:1991rp}. 
While this approach has been very successful, the future improvement
of observational data will necessitate knowing the theoretical error
arising from the PT formalism assumptions.  In fact, the late-time
contribution to nongaussianities from PT is orders of magnitude
larger than that expected from primordial nongaussianity 
\cite{Scoccimarro:2003wn}, and even relatively small corrections
to the PT ansatz may swamp the signal we are interested in.

Experimental limits on non-Gaussianities are generically given in terms
of a scalar variable $f_{\rm NL}$ \cite{Komatsu:2003fd, Babich:2004gb} which 
in the so-called ``local'' ansatz parameterizes the deviation of
the Newtonian potential from a gaussian variable $\Phi_g$ as
\be
\Phi &=& \Phi_g + f_{\rm NL} \left( \Phi_g^2 - \langle \Phi_g^2 \rangle
\right) 
\ee
There are other possible shapes from models of inflation,
but we will use this one for comparison.
We find that radiation effects on the LSS are comparable in size to
$f_{\rm NL} \sim 4$.  Current LSS experimental limits are still consistent
with a gaussian distribution and have 95\% confidence limits of 
$-29  < f_{\rm NL} < 69$ \cite{Slosar:2008hx}.  So far, they are nicely consistent with the 
limits from the CMB \cite{Smith:2009jr}, which, when combined together, give the constraint $-1<f_{\rm NL}<63$ at 95\% confidence level.

\section{\label{sec:estimates} Estimates  }

\subsection{PT Formalism}

We are ultimately interested in calculating the difference between
the three-point function from the PT formalism and from a universe
with a radiation component, and comparing this difference to 
the three-point function from primordial nongaussianities.  Consider
first what kind of
effects we expect for small vs. large wavelength modes.
For very large
scales that cross the horizon deep in the matter dominated region,
we do not expect any growth in $\delta$ before the matter-dominated
region begins and thus we can essentially treat the entire history
of the universe as matter dominated.  However, for short wavelength modes,
growth has already begun deep in the radiation dominated region and
it is not immediately clear whether this can give a large correction or
not to the distribution of perturbations. 
 The perturbations are very small during the
radiation dominated era, and this implies that non-gaussianities
generated during this region will be very small.  That is,
non-gaussianities from gravitational instability are generated
because the first order fluctuations are a source for the
second order fluctuations:
\be
\dot{\delta}^{(2)} &\sim& \delta^{(1)2}\ ,
\ee
and thus this generates non-gaussianity dominantly at recent times.
In this section, we will estimate the effect of radiation
by considering how a large distortion in $\dot{\delta}$ during the
radiation era affects the three-point function at late times.

We first recall a few results. The method of PT is essentially to perform
an expansion in the scale factor $a$.  At second order,
\be
\delta(a,k) &=& \delta_1(k) a + \half \delta_2(k) a^2\ , \\
\delta_2(k) &=& \int {d^3 q_1 d^3 q_2 \over (2\pi)^3} \delta(k-q_1 - q_2)
   F_2(q_1,q_2) \delta_1(q_1) \delta_1(q_2)\ . \nn\\ 
\ee
In Newtonian gravity with matter only and no decaying mode, one
obtains~\footnote{The reader familiar with the PT literature should
note the additional factor of $\half$ in the expression for $F_2$.
This is a consequence of the additional factor of $\half$ in our
convention for $\delta^{(2)}$. }  
\be
\half F_2(q_1,q_2) &=& {5 \over 7} +{\hat{q}_1 \cdot \hat{q}_2 \over 2}
 \left( { q_1 \over q_2} + { q_2 \over q_1} \right) + 
  {2 \over 7} (\hat{q}_1 \cdot \hat{q}_2)^2 \ . 
\label{eq:ptkernel}
\ee
Thus, as a function of $a$, the three-point function is
$\langle \delta_{k_1} \delta_{k_2} \delta_{k_3} \rangle (a) 
  \sim  a^4 P^2 F_2 + a^3 B_1 (k_1,k_2,k_3)$,
where $P (B_1)$ is the two-point (three-point) function of 
$\delta_1$,~\footnote{Power spectra for fluctuations other than $\delta$
will be denoted with a subscript to indicate the fluctuation, e.g.
$P_\phi(k)$ for $\Phi$ fluctuations.} and thus $B_1$ encodes
the three-point function initial condition. Modifying the evolution
history up to a scale factor around $a_{\rm eq}$ will modify
$B_1$ proportionally by $a_{\rm eq}$.  For instance, suppressing
all growth of nongaussianity in this simple approximation up until
$a_{\rm eq}$, so that $\langle \delta_{k_1} \delta_{k_2} \delta_{k_3} \rangle
(a_{\rm eq}) =0$, imposes 
\be
B_1 \sim - a_{\rm eq} P^2 F_2\ .
\label{eq:BGappx}
\ee
We therefore
ought to expect corrections to the three-point function today
proportional to $a_{\rm eq}$.  

Such a small correction matters because at late times and short wavelengths, 
the nongaussianity from nonlinear growth (i.e. from $F_2$)
swamps the contribution from initial conditions. Within
linear theory, $\delta_k(a) = M_k(a) \Phi_k^{\rm prim}$,
\be
M_k(a) &=& -{3 \over 5} {k^2 T(k) \over \Omega_m H_0^2 } D_1(a)\ ,
\ee
where $D_1(a)$ is the growing mode for $\delta$ and $T(k)$
is the transfer function.  Under the local ansatz 
$\Phi^{\rm prim} = \Phi_g + f_{\rm NL} \Phi_g^2$, this
leads to a dark matter three-point function proportional to $f_{\rm NL}$:
\be
B_L (k_1,k_2,k_3) &=& M_{k_1} M_{k_2} M_{k_3}  2 f_{\rm NL}
   \left( P_\Phi(k_1) P_\Phi(k_2) + \textrm{cyc.} \right) \ .
\ee

The full three-point function is then approximately the sum of the 
contribution $B_L$ from linear growth and the contribution $B_G$ from
non-linear growth. 
For the sake of this approximation, we will take isoceles triangle
configurations $|k_1|=|k_2| \equiv y |k_3|$, in which case 
$\half F_2(k_1,k_2)={1 + 3 y^2 \over 14 y^4}$ and 
$\half F_2(k_1,k_3)=\half F_2(k_2,k_3) = {1 \over 28} (13 - {5 \over y^2})$. Note
that $y\ge \half$.  
We will also ignore the scalar tilt $n_s$, so $P_\Phi(k) \propto k^{-3}$.  
Now compare this to a correction in the nonlinear bispectrum
of the size we expect from modifying the model at scale factor 
$a_{\rm eq}$, i.e. $ \delta B_G \approx a_{\rm eq} F_2(k_1,k_2) P(k_1)P(k_2) 
+ \textrm{cycl.}$:
\be
{ \delta B_G \over B_L } &\approx&
  {a_{\rm eq} M_{k_3} \over 2 f_{\rm NL} } 
   \left( {2y^3 F_2(k_1, k_3) + F_2(k_1,k_2) \left( {M_{k_1} \over M_{k_3}} \right)^2  
 \over  2y^3 +1 }\right)   \ .
\ee
We can simplify this in various limits.  The transfer functions
$T(k)$ simplify at large and at small $k$ in a flat (matter + radiation) 
universe:
\be
&&T(k \ll k_{\rm eq}) = 1\ , \\
&&T(k \gtrsim 15 k_{\rm eq} ) = {45 \over 2} { \Omega_m H_0^2 
\over k^2 a_{\rm eq}} \log(k/6 k_{\rm eq})\ .
\ee

In the limit of squeezed triangles $y \gg 1$, we arrive at the following
estimate
\be
{ \delta B_G \over B_L } &\approx&
   {a_{\rm eq} M_{k_3} \over f_{\rm NL} } { 13 \over 28} 
  =  \left\{ \begin{array} {ll}
   {-0.56  \over f_{\rm NL} \Omega_m } \left( {k_3 \over k_{\rm eq}} \right)^2
 & k_3 \ll k_{\rm eq} \\
  {-6.3 \over f_{\rm NL} } \log (k_3/6k_{\rm eq}) &
 k_3 \gg k_{\rm eq} \end{array} \right\}\ ,  \qquad y \gg 1\ .
\ee
Note that all dependence on the shorter wavelength modes
$k_1,k_2$ drops out.  Thus, even if $k_1,k_2$ enter the horizon
deep in the radiation era, the 3-pt function is unaffected by
the radiation corrections as long as $k_3$ enters during
matter dominance (MD), where $\delta B_G$ is suppressed by $(k_3/k_{\rm eq})^2$;
however, the 3-pt function is affected if $k_3$ enters during the
radiation era.

We next turn to the limit of equilateral triangles, $y=1, k\equiv k_i$:
\be
{\delta B_G  \over B_L } &\approx& {a_{\rm eq} M_k \over f_{\rm NL}} {2 \over 7}
  =  \left\{ \begin{array}{ll}
   {-0.34 \over f_{\rm NL} \Omega_m} \left({ k \over k_{\rm eq}} \right)^2
& k \ll k_{\rm eq} \\
   {-3.9 \over f_{\rm NL} } \log (k/6 k_{\rm eq}) & k \gg k_{\rm eq}
\end{array}  \right\}\ , \qquad y=1\ .
\ee
In other words, modifying the matter content of the universe
at $a_{\rm eq}$ near or after matter-radiation equality alters the
bispectrum roughly corresponding to $f_{\rm NL} \sim \textrm{few}$ for
modes that entered the horizon somewhat before matter-radiation
equality. 
We will need a more precise calculation of nonlinear growth
near matter-domination equality in order to predict
the bispectrum at order $f_{\rm NL} \sim few$.

\subsection{Radiation Effect in Newtonian Gravity}
\label{sec:radnewt}

By the previous arguments, we expect that the dominant contribution
from radiation will come from the era near the transition between matter
and radiation, because at this time, $\delta$ is as large as possible
before matter dominates.  Short wavelength modes, which by previous
arguments will get the largest contribution, will therefore already be
inside the horizon at that point and we should be able to approximate
the effect on them using Newtonian gravity.  In this limit, we will
be able to obtain an analytic approximation.

The equations of motion for the dark matter perturbations in Newtonian
gravity are (e.g. \cite{Bernardeau:2001qr})
\be
&& \delta'_k + i k V_k = -\left( 1 + {q_1 \cdot q_2  \over q_2^2} 
\right) \delta_{q_1} i q_2 V_{q_2}  
\label{eq:newtrad1}\ , \\
&& ik V'_k + \CH i k V_k + {3 \over 2} \CH^2 \Omega_m(\eta) \delta_k = \half
k^2 \hat{q}_1 \cdot \hat{q}_2 V_{q_1} V_{q_2}\ .
\label{eq:newtrad2}
\ee
Primes denote derivatives with respect to $\eta$. 
The behavior of the background
is that of a universe with matter and radiation with no cosmological
constant, which can be written
in terms of $a_{\rm eq} = { \Omega_r \over \Omega_m}$  as \cite{Seljak:1994yz}
\be
&&{a \over a_{\rm eq} } = 2 \alpha \eta + \alpha^2 \eta^2\ ,
\qquad 
\alpha = { H_0 \over 2\sqrt{a_{\rm eq} (1+a_{\rm eq})} }
  = {\sqrt{2}-1 \over \eta_{\rm eq}}\ , 
\label{eq:background}
\ee
and $\Omega_m(\eta) = {\rho_m \over \rho_{\rm tot}} = {1 \over 1 + a_{\rm eq}/a}$.
It is straightforward to solve equations (\ref{eq:newtrad1},\ref{eq:newtrad2})
at linear order.  The solution~\footnote{We take the initial conditions at
arbitrarily small $\eta$, so the decaying mode is projected out and
the solution is the growing mode.} normalized so that $\delta^{(1)}=1$ today is
\be
&&\delta_g^{(1)}(\eta) = {2+ 3 \alpha \eta (2 + \alpha \eta)
\over 2 + 3 \alpha \eta_0 (2 + \alpha \eta_0)} \ , \qquad
i k V_g^{(1)}(\eta) = -{ 6 \alpha (1 + \alpha \eta) \over
2 + 3 \alpha \eta_0 (2 + \alpha \eta_0)}\ .
\ee
Taking the first order solutions to contain only a growing mode, 
we obtain an analytic solution for the second order modes, which for
brevity we do not write out.  Instead we expand in $1/\alpha$, and
obtain the PT result equation (\ref{eq:ptkernel}) plus corrections.
The leading correction is enhanced by $\log (a_{\rm eq})$:
\be
\half F_{2,rad} &=& \half F_{2,PT} - \half a_{\rm eq} \log \left(  
a^{-1}_{\rm eq}  \right)
 { 1 \over 35 q_1^2 q_2^2} \left( k^2 - (q_1 + q_2)^2 \right)
\left( k^2 -(q_1 -q_2)^2\right) + \dots \nn\\
&=& \half F_{2,PT} + \half a_{\rm eq} \log \left(  
a^{-1}_{\rm eq}  \right)
 { 4 \over 35} \left(1- (\hat{q}_1 \cdot \hat{q}_2)^2 \right) + \dots\ . \nn\\
\label{eq:f2rad}
\ee
For example, on equilateral configurations
(i.e. $k=q_1=q_2$), we have 
$ F_{2,rad} - F_{2,PT} 
\approx a_{\rm eq} {3 \over 20} \log (a_{\rm eq}^{-1}) F_{2, PT}$, 
which is $\approx  1.2 a_{\rm eq} F_{2,PT}$ for $a_{\rm eq} = 3 \times 10^{-4}$.  
As in the case without radiation, the kernel $F_{2,rad}$ is invariant
under rescalings $(k,q_1,q_2) \rightarrow \lambda (k,q_1,q_2)$,
and all scale-dependence of $\delta^{(2)}$  
comes from the scale-dependence of $\delta^{(1)}$.  Note
that the leading term vanishes on ``squashed'' triangle configurations
$k = |q_1 \pm q_2|$, and in this case the subleading, non-log-enhanced
contribution dominates.  However, the non-log-enhanced contribution
is more sensitive to the behavior of the linear solutions during the
radiation era, and in order to get an accurate estimate valid at large
$k$ we would have to include some effects we have neglected.

In the above approximation, we included radiation only in the background
dependence, i.e. in $\CH(\eta)$. This approximation is exact in the limit
$k \gg k_{\rm eq}$. A more careful treatment would
include the effect of the radiation density perturbations as well,
which would enhance the evolution of the Newtonian potential $\Phi$
at early times.  One might then obtain an analytic approximation
by including the backreaction of only the radiation on the Newtonian potential
at early times and only the matter at later times, and matching the
two solutions in an intermediate regime.  One would expect in this way to see
an enhancement of the final perturbation size, 
similar to results from the analogous
method in linear perturbation theory
\cite{Dodelson:2003ft}.  However, we instead will now turn to a numeric
calculation that will give the fluctuations at second order
for all wavelengths.

\section{Method \label{sec:method}}

In this section we will describe our method for calculating
the effects of radiation on the dark matter density three-point
function at late times, based on an expansion at second order in the
fluctuations.  That is, we split all perturbations $\CO = \CO^{(1)}
+ \half \CO^{(2)}$ into
a first order piece $\CO^{(1)}$ and a second order piece $\CO^{(2)}$.
For closely related work on second order
corrections to cosmological perturbations see e.g. \cite{Bartolo:2003bz,Bartolo:2005kv,Matarrese:1997ay,Bartolo:2006fj,Bartolo:2007ax,Boubekeur:2008kn}.
Since we are interested in following the dark matter distribution, 
 we can approximate the baryon-photon fluid as a fluid with 
$w={ 1 \over 3}$, and otherwise neglect $\Omega_b$  for simplicity, taking a universe of two perfect fluids
composed respectively of collisionless
matter and radiation. In this approach, the second order perturbations
are sourced due to nonlinear interactions by the first order ones, and 
thus are quadratic in them, $\CO^{(2)}
 \sim \CO^{(1)} \CO^{(1)}$. This leads to a nonvanishing three-point function
even for purely gaussian initial conditions, and this contribution
should be removed in order to reconstruct the primordial nongaussianity.
In particular, we are interested in the kernel $F_2$ in Fourier space:
\be
&& \half \delta^{(2)}_{\vec{k}}(\eta) = \half \int {d^3 q_1 d^3 q_2 \over (2\pi)^3} 
\delta(\vec{k} - \vec{q}_1 - \vec{q}_2) F_2(\vec{q}_1,\vec{q}_2;\eta)
\delta^{(1)}_{\vec{q}_1}(\eta_0) \delta^{(1)}_{\vec{q}_2}(\eta_0)\ ,
\ee
where $\eta$ is conformal time and we choose $\eta_0=\eta_{\rm today}$
in order to make contact with existing literature. We compute $F_2$
in a straightforward manner for any $q_1,q_2$ by inputting the first
order solutions into the equations of motion as sources and solving
numerically for the second order terms.  This procedure could be carried
out iteratively to higher orders as well by inputting the first and
second order solutions as sources at third order, etc.  

We also may neglect all vector and tensor degrees of freedom.
 The reason is that vector
and tensor degrees of freedom vanish at first order in the initial
scalar perturbations, arising only at second order.  Therefore,
we may decompose any equation with a vector index into its
scalar $\CO_s$ and vector $\CO_v^i$ piece.  In Fourier space, the most general
equation at second order is
\be
(a q_1^i+b q_2^i) \CO_s^{(1)}(q_1) \CO_s^{(1)}(q_2)
   + k^i \CO_s^{(2)}(k) + \CO_v^{i (2)}(k)&=& 0\ ,
\ee
where $k^i \CO_v^i(k) =0$. By contracting this equation with $k^i$,
we therefore project out all vector modes and
 obtain a second order equation for just the scalar modes.
A similar argument applies to the tensor fluctuations that arise from
primordial scalar fluctuations.  In addition,
tensor modes have their own primordial fluctuations, whose size is
typically suppressed with respect to the scalar fluctuations in models 
of inflation. More important, their
two-point function with the scalar modes necessarily vanishes by
rotational invariance, and thus
they give no contribution to the matter three-point function.  We may
therefore neglect them as well.

The computation is simplest in conformal Newtonian gauge, where the metric
is diagonal.  Specifically,  we will take
\be
ds^2 &=& a^2 \left( -e^{2 \Psi} d\eta^2 + 2 \omega_i dx^i d\eta
  + (e^{-2 \Phi} \delta_{ij} + \chi_{ij} ) dx^i dx^j \right) \ ,\\
\omega^i_{,i} &=& 0, \qquad \chi_{ij,j} =0 \ ,
\ee
so that $\omega_i, \chi_{ij}$ contain no scalar modes.  We may therefore
neglect $\omega$ and $\chi$ completely for following the
scalar modes at second order.

The energy-momentum tensor is that for a fluid of matter,
which is pressureless up to negligible $\CO({T \over m})$ corrections, and a
fluid of radiation $w=1/3$:
\be
&&T^\mu_{\ \nu,m} =  \rho_m u_m^\mu u_{\nu,m} \ ,\qquad
T^\mu_{\ \nu,r} = \rho_r \left( {4 \over 3} u_r^\mu u_{\nu,r} + {1 \over 3}
\delta^\mu_\nu \right)\ , \qquad
u^\mu_{m,r} = \left( {e^{-\Psi} \over a} (1+{v_{m,r}^2 \over 2} ),
{e^\Phi \over a} v^i_{m,r} \right)\ . \nn\\
\label{eq:Tmunu}
\ee

We will denote the longitudinal piece of velocity as $v^i = \hat{k}^i V$.
We obtain the equations of motions from Einstein's
equation $G^\mu_{\ \nu} = 8 \pi G_{\rm N} T^\mu_{\ \nu}$
and the conservation equations
$\nabla_\mu T^\mu_{\ \nu,m,r}=0, \nabla_\mu(\rho_m u^\mu_m) =0$, 
which is valid because matter and radiation interact only
gravitationally.  

It is straightforward to derive the Einstein tensor
exactly in conformal Newtonian gauge with only scalar
modes.  We write it down for reference in appendix \ref{app:GRMD}. 
There are two components of the Einstein equation that will be
useful to us.  The first is the time-time component,
$a^2 G^0_{\ 0} =  3 \CH^2 T^0_{\ 0}/\bar{\rho}$, which in Fourier space is
\be
&& k^2 \Phi_k^{(2)} + 3 \CH \Phi_k'^{(2)} + 3 \CH^2 \Psi_k^{(2)}  
   +{3 \over 2} \CH^2
 \left( {\bar{\rho}_m \over \bar{\rho}} \delta^{(2)}_m
  + {\bar{\rho}_r \over \bar{\rho}} \delta^{(2)}_r \right)  = -S_3 \equiv 
\label{eq:e3}\\
&& \qquad \left( q_1 \cdot q_2 -4 q_2^2+6\CH^2 \right)
   \Phi_{q_1}^{(1)} \Phi_{q_2}^{(1)}  
  + 3 \Phi_{q_1}'^{(1)} \Phi_{q_2}'^{(1)} 
  + 12 \CH \Phi_{q_1}^{(1)} \Phi_{q_2}'^{(1)} \nn\\
&& \qquad  - 3 \CH^2 \hat{q}_1 \cdot \hat{q}_2
 \left({\bar{\rho}_m \over \bar{\rho}}  V_{q_1,m} V_{q_2,m}
  +{4 \over 3} {\bar{\rho}_r \over \bar{\rho}} V_{q_1,r} V_{q_2,r} \right)\ . \nn
\ee
Primes denote $\eta$ derivatives, and $\CH = {a' \over a}$. 
In the above equation, an integral
 $(2\pi)^{-3}\int d^3 q_1 d^3 q_2 \delta(\vec{k} -
  \vec{q}_1 - \vec{q}_2)$
is implied over terms quadratic in the
first order perturbations.  In general, we mean for such an integral to
be implicit in any equation in Fourier space with terms quadratic
 in the first order perturbations.  The second useful component
is the shear piece, specifically
$(\hat{k}^i \hat{k}^j - {1 \over 3} \delta_{ij}) G^i_{\ j}$, which projects
out the $\delta_{ij}$ piece in $G^i_{\ j}$. 
The resulting equation of motion is
\be
&& {1 \over 3} k^2 ( -\Phi^{(2)} + \Psi^{(2)} )  = 
   {1 \over 3} k^2 S_4(k,\eta) \equiv 
\label{eq:e4} \\
&& \qquad   \left(
      (-2(\hat{k} \cdot q_1)(\hat{k} \cdot q_2) + {2 \over 3} q_1 \cdot q_2)
          \Phi^{(1)}_{q_1} \Phi^{(1)}_{q_2} \right) \nn\\
&& \qquad
  + 3 \CH^2 ( ( \hat{k} \cdot \hat{q}_1) (\hat{k} \cdot \hat{q}_2)
  - {1 \over 3} \hat{q}_1 \cdot \hat{q}_2 ) 
 \left({\bar{\rho}_m \over \bar{\rho}}  V_{q_1,m} V_{q_2,m}
  +{4 \over 3} {\bar{\rho}_r \over \bar{\rho}} V_{q_1,r} V_{q_2,r} \right)\ .
 \nn
\ee 
We now turn to the conservation equations for matter.
The first comes from $\nabla_\mu (\rho_m u^\mu_m) =0$, the second from
$\nabla_\mu T^\mu_{\ i,m}=0$:
\be
&&\delta'^{(2)}_m + i k V^{(2)}_m - 3 \Phi'^{(2)} = S_1(k,\eta) \equiv 
\label{eq:e1} \\
  && \qquad 
   2 \delta_{q_1,m}^{(1)} \delta_{q_2,m}'^{(1)}  - 2i (q_1 \cdot \hat{q}_2)
  \delta_{q_1,m}^{(1)} V_{q_2}^{(1)} 
   - 4 i q_2 \Phi_{q_1}^{(1)} V_{q_2,m}^{(1)} +2i (\hat{q}_1 \cdot q_2)
V_{q_1,m}^{(1)} \Phi_{q_2}^{(1)}
  -2 (\hat{q}_1 \cdot \hat{q}_2) V_{q_1,m}^{(1)} V_{q_2,m}'^{(1)}
 \nn\ ,\\
&&V'^{(2)}_m + \CH V^{(2)}_m + i k \Psi^{(2)} = S_2(k,\eta)\equiv 
\label{eq:e2}\\
&& \qquad
   2 (\hat{k} \cdot \hat{q}_2) \Phi'^{(1)}_{q_1} V_{q_2,m}^{(1)}
  -ik (\hat{q}_1 \cdot \hat{q}_2) V_{q_1,m}^{(1)}
  V_{q_2,m}^{(1)} -2 ik \Phi_{q_1}^{(1)} \Phi_{q_2}^{(1)}\ .
\nn
\ee
where we have simplified the sources $S_i$ by imposing the first-order equations
of motion and symmetrizing in $q_1,q_2$ when convenient.

We take the two radiation equations of motion from $\nabla_\mu T^\mu_{\ i,r}=0$
and $\nabla_\mu T^\mu_{\ 0,r}=0$:
\be
&& \delta'^{(2)}_r - 4 \Phi'^{(2)} + {4 i k \over 3} V^{(2)}_r = S_6  \equiv
\label{eq:e6} \\
&& \qquad -{4 \over 3} i \hat{q}_1 \cdot q_2 V^{(1)}_{q_1,r} \delta^{(1)}_{q_2,r}
+ {16 \over 3}i \hat{q}_1 \cdot q_2 V^{(1)}_{q_1,r} \Phi^{(1)}_{q_2} 
 - {16 \over 3} i q_2 \Phi^{(1)}_{q_1} V^{(1)}_{q_2,r} +2
 \delta^{(1)}_{q_1,r}
\delta'^{(1)}_{q_2,r}\ ,\nn\\
&&V'^{(2)}_r + {i k \over 4} \delta^{(2)}_r + i k \Psi^{(2)} = S_7 \equiv
\label{eq:e7} \\
&& \qquad {i \over 3 k} \left( 2q_1 q_2 + (q_1^2 + q_2^2-3k^2) \hat{q}_1 \cdot
\hat{q}_2 \right) V^{(1)}_{q_1,r} V^{(1)}_{q_2,r} 
  + {i k \over 4} \delta^{(1)}_{q_1,r} \delta^{(1)}_{q_2,r}
 + 4 \hat{q}_2 \cdot \hat{k} \Phi^{(1)}_{q_1} V'^{(1)}_{q_2,r} \ .\nn
\ee

  The behavior of the background
is that of a universe with matter and radiation, as in equation
(\ref{eq:background}).
In addition to the above equations of motion, we require the 
 equations for the first order perturbations.
The equations of motion for the first order perturbations are identical in
form to those for the second order solutions, with the sources
$S_i$ set to zero.  We solve these numerically.

The initial conditions for the fluctuations
depends on physics
before they reenter the horizon.  There is both a contribution
from before a mode exits the horizon during inflation that depends
on the inflationary model as well as a possible contribution
due to light degrees of freedom that perturb the reheating surface.  
Outside the horizon,
it is useful to work with the metric of scalar modes
can be set by comparing with $\zeta$ gauge, which in the absence of
spatial gradients
is  defined at nonlinear order by
\be
ds^2 &=& a^2(\eta) \left( -d \eta^2 + e^{2 \zeta} dx^i dx^i \right)\ .
\label{eq:zetagauge}
\ee
For the minimal model of inflation, $\zeta$ is initially an approximately
gaussian field with a three-point function that is suppressed
by a slow-roll parameter \cite{Maldacena:2002vr,Acquaviva:2002ud,Salopek:1990jq}. 
 We therefore choose 
$\zeta^{(2)} =0$ initially, in order to separate out the
effects due to nonlinear growth at late times. This condition in turn
sets the initial conditions on most of the remaining perturbations
through the equations of motion.  Let us first consider how
it sets $\Psi^{(2)}$ and $\Phi^{(2)}$ initially.  $\Psi^{(2)}$ and
$\Phi^{(2)}$ are related to each other through the constraint equation 
(\ref{eq:e4}), which enforces
\be
\Psi^{(2)} &=& \Phi^{(2)} + S_4\ .
\ee
At nonlinear order, transforming from $\zeta$ gauge (\ref{eq:zetagauge})
to Newtonian gauge  shows that 
$\zeta = - \Phi - \half \Psi$ during the radiation era in the absence
of spatial gradients.  Thus, at $a \ll a_{\rm eq}, k \eta \ll 1$
\be
&& \Phi = -{2 \over 3} \zeta -{1 \over 6} S_4\ ,
\ee
and therefore, we have
$\half \Phi^{(2)} = -{1 \over 6} S_4$ initially.

The initial conditions for the perturbations therefore 
depend on the behavior of the source
terms at early times $a \ll a_{\rm eq}$ during the radiation
era and long wavelengths $k \eta \ll 1$, because as above the equations
of motion each contain perturbations without time derivatives.  At early times,
these perturbations contain only a growing mode, which is usually constant
outside the horizon, and the equations of motion turn into an algebraic
relations between the perturbations and the sources.  This is 
also what happens in the linear theory, except that the source terms
$S_i$ are not present in that case.  In the long-wavelength early-time limit, 
we will see that we need to use only
 $S_3 \rightarrow -6 \CH^2 \Phi_{q_1}^{(1)}
\Phi_{q_2}^{(1)}$ and $S_4 \rightarrow 
-{9 \over k^2} \left( \hat{k} \cdot q_1 \hat{k} \cdot q_2 -
{1 \over 3} q_1 \cdot q_2 \right) \Phi^{(1)}_{q_1} \Phi^{(1)}_{q_2}$
(using $i V_r^{(1)} = \half k \eta \Phi^{(1)}$ initially). 
Equation (\ref{eq:e3}) and (\ref{eq:e4}) together imply the 
initial condition for $\delta_r^{(2)}$:
\be
&&\delta_r^{(2)} 
 = - {2 \over 3 \CH^2} S_3 - 2 S_4- 2 \Phi^{(2)}  \ .
\label{eq:secondorderinit}
\ee
The second order
velocities $V^{(2)}_m, V^{(2)}_r$, start off negligible at early times;
if needed, they can be read off of equations (\ref{eq:e2},\ref{eq:e7})
by taking $V^{(2)} \propto \eta$ at small $\eta$.  

Finally, we need to set the initial condition for $\delta_m^{(2)}$.
It does not appear in the equations of motion at early times, because
it always has a time derivative acting on it or else is multiplied
by $(\bar{\rho}_m / \bar{\rho})$, which vanishes in the infinite past.
We will assume adiabatic initial conditions, where there is just a
single light scalar degree of freedom outside the horizon (i.e.
no entropic modes).  In this case, we may set the initial condition
on $\delta_m^{(2)}$ by the fact that all scalar fluctuations arise
from the time-shift of their background.
More precisely, for adiabatic initial conditions, all scalar fluctuations 
outside the horizon
arise from a single scalar fluctuation, which may be parameterized
as the pion $\delta t$ for the spontaneously broken time translations \cite{us}.
Thus, for either matter or radiation, $\bar{\rho} + \delta \rho =
\bar{\rho}(t+ \delta t) = \bar{\rho}(t) + \delta t \dot{\bar{\rho}}(t) + \half
\delta t^2\ddot{\bar{\rho}}(t)+\dots$.
At first order, $\delta t^{(1)} = {\delta \rho \over \dot{\rho}}$ is therefore
equal for matter and radiation, and this implies the usual adiabatic relation
$\delta_m = {3 \over 4}\delta_r$ initially.  At second order, 
one must equate $\delta t^{(2)} = {\delta \rho^{(2)} \over \dot{\rho}} -
{\ddot\rho \over \dot{\rho}} \left( \delta \rho^{(1)} \over \dot{\rho}
\right)^2
$, and therefore the initial condition for $\delta_m^{(2)}$ is
\be
\delta_m^{(2)} &=& {3 \over 4} \delta_r^{(2)} - {3 \over 16}
\delta_r^{(1)2}\ .
\ee

In principle, the equations of motion 
(\ref{eq:e3}-\ref{eq:e7})
and the initial
conditions are all that is needed to find the second order kernel
on any scale.  In practice, however, we also must specify the gauge
in which $\delta_m^{(2)}$ is actually measured in observations.  While
it is straightforward now to calculate $\delta_m^{(2)}$ in conformal
Newtonian gauge, it is unlikely that this is the quantity we are
interested in.  We shall not determine in this paper which
is the correct gauge for $\delta_m$ corresponding to measurements
(or, put differently, what is the correct gauge-invariant observable
measured in surveys).
Rather, we shall treat gauge dependence as an additional source
of uncertainty. What this means requires some care, as one can
always choose the gauge $\delta_m = 0$,
where the matter density provides the clock for the coordinate system.
We shall avoid such extreme gauge choices and instead consider the
change in $\delta^{(2)}$ going from Newtonian to synchronous gauge.
On subhorizon scales $k/\CH \gg 1$, perturbations become increasingly
local and insensitive to the gauge choice, and $\delta$ from one
gauge to the next changes as $\sim {\CH^2 \over k^2} \delta $. 
For example, under a time diffeomorphism $\eta \rightarrow \eta+ \alpha$,
$\Psi$ changes at first order by $\Psi \rightarrow \Psi -
\CH \alpha - \alpha'$, and $\delta $ changes at second order
by $\delta \rightarrow \delta + 3 \CH \alpha (1+ \delta) + \dots$.  
Thus inside the horizon during the matter era,
a gauge transformation that changes the metric $\Psi$ by ${\cal O}(1)$
has $\CH \alpha \sim \Psi$, and consequently $\delta^{(2)}
\sim 3 \Psi^{(1)} \delta^{(1)} \sim 3{(\delta^{(1)})^2 \over M_k}
\sim 5 { \CH_0^2 \over k^2} (\delta^{(1)})^2$.  Thus the physical significance of
contributions to $F_2$ that behave parametrically as $\CH^2/k^2$ is unclear,
and much additional work is required to understand these contributions.

\section{Results \label{sec:results}}

\subsection{Numeric Results}

The second order kernel $F_2$ is a function of two three-momenta
$\vec{q}_1$ and $\vec{q}_2$ and is thus naively a function of
six variables.  However, due to rotational invariance, it
is in fact a function of only three variables, which we will
choose to be $k, \ x_1 \equiv {q_1 \over k},$ and $x_2 \equiv {q_2 \over k}$.
In the limit of short wavelengths and negligible radiation, 
$F_2$ approaches the PT result of equation (\ref{eq:ptkernel}),
which is scale invariant and thus depends only on $x_1$ and $x_2$. Note
that $1+ x_2 > x_1 > 1-x_2$ by momentum conservation and we may take
$x_1 > x_2$ without loss of generality by symmetry.  
The kernel $F_2(q_1,q_2)$ enters the three-point function
through
\be
\langle \delta_{k_1} \delta_{k_2} \delta_{k_3} \rangle &=&
\delta\left( \sum_i \vec{k}_i \right) (2\pi)^3
\left(  F_2(k_1,k_2)P_{k_1}P_{k_2} + \textrm{cyc.} \right)
 \nn\\
&=& B_G(k_1, k_2, k_3) \times (2\pi)^3\delta\left( \sum_k \vec{k}_i \right)
\ ,
\ee
and in the three-point function one may further take $k_1 > k_2 > k_3$
without loss of generality. 
It is convenient to introduce the reduced three-point function
\be
Q(k_1,k_2,k_3) &=& {B_G(k_1,k_2,k_3) \over
   P(k_1)P(k_2) + P(k_2)P(k_3)+ P(k_1)P(k_3) }\ .
\ee
$Q(k_1,k_2,k_3)$ is identical to $F_2(k,k)$ on equilateral triangles. 
Unlike $F_2$, though, it is symmetric under permutations of $k_1,k_2,k_3$,
and thus takes into account some cancellations in the three-point
function between $F_2(k_1,k_2)$ and its permutations.

\begin{figure}[t!]
\begin{center}
\includegraphics[width=0.48\textwidth]{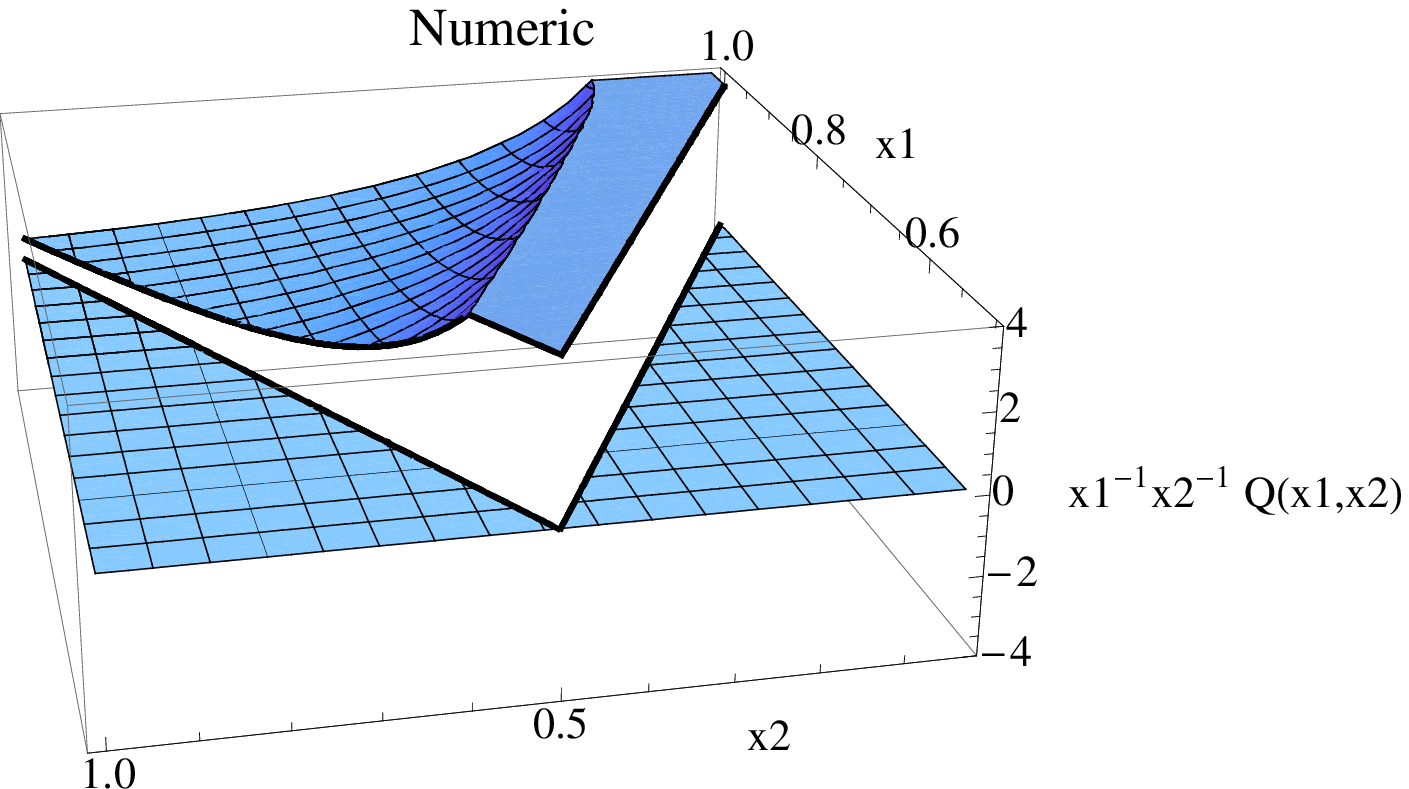}
\includegraphics[width=0.48\textwidth]{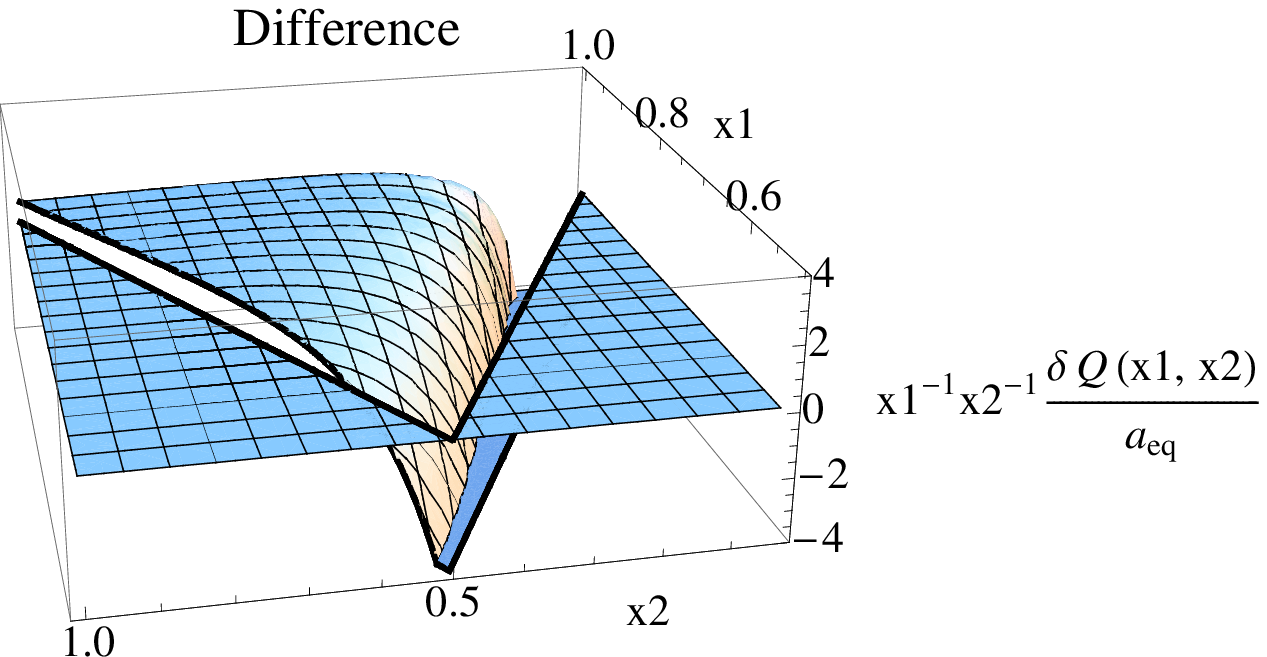}
\caption{The kernel $x_1^{-1} x_2^{-1} Q(x_1, x_2)$ for $k$ fixed at
$k=11 k_{\rm eq}$.  The left plot is from our numeric computation, and
is indistinguishable by eye from the PT limit given in equation (\ref{eq:ptkernel}). The right plot is the difference between the numeric
result and the PT result, rescaled by $a_{\rm eq}$, because
by equation (\ref{eq:BGappx}), $a_{\rm eq}$ controls the size of the
radiation correction. }
\label{fig:ptf2}
\end{center}
\end{figure}

\begin{figure}[ht!]
\begin{center}
\includegraphics[width=0.68\textwidth]{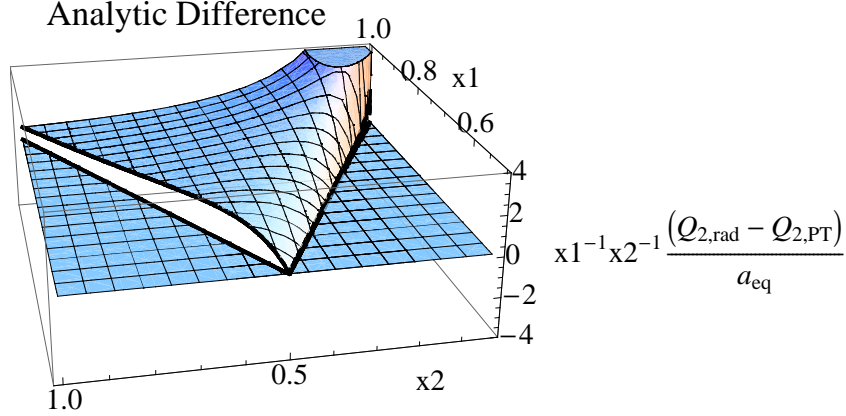}
\caption{The log-enhanced difference between the PT kernel and the kernel from
using Newtonian gravity with a background radiation component.
The analytic result is given in equation (\ref{eq:f2rad}). }
\label{fig:df2rad}
\end{center}
\end{figure}

A plot of $x_1^{-1} x_2^{-1} Q(x_1,x_2)$ from our numeric computation
with $k$ fixed at $11 \ k_{\rm eq}$ is plotted in Figure \ref{fig:ptf2},
along with a plot of the difference between our numeric result
and the PT result~\footnote{
We define the ``difference'' $\delta Q$ as the change from corrections
to $F_2$, and not from corrections to the power spectrum $P(k)$.
 Explicitly,
\be
\delta Q &\equiv& 
{ \delta F_2(k_1,k_2) P(k_1) P(k_2) + \delta F_2(k_2,k_3)
P(k_2) P(k_3) + \delta F_2 (k_3,k_1) P(k_3) P(k_1) \over
P(k_1) P(k_2) + P(k_2) P(k_3) + P(k_3)P(k_1) }\ ,  \nn\\
\delta F_2 (q_1,q_2) &\equiv& F_{2,exact} - F_{2,PT}\ .
\ee
}. The observed value of $k_{\rm eq}$ is $0.013 h {\rm Mpc}^{-1}$.
At $k=11 \ k_{\rm eq}$, the difference is very close
to that obtained in Newtonian gravity with radiation, equation
 (\ref{eq:f2rad}), shown in Figure \ref{fig:df2rad}.  This
indicates that ${k \over k_{\rm eq}} \gtrsim 10$ is already large
enough that the approximation used in section \ref{sec:radnewt}
is good.


The result from our numeric method takes into account not only radiation
corrections but also corrections from general relativity (GR), the leading
contributions of which are post-Newtonian (PN) corrections.
The scale-dependence of these two effects is completely different.
GR effects are important on larger scales and behave parametrically
as $(\CH/k)^2$ compared to the PT kernel.  We can partially isolate such 
effects by using our numeric method with $\Omega_r=0$ (equivalently,
$a_{\rm eq}=0$), and taking the initial conditions appropriate for a matter-only
universe.  In fact, in this case the second order perturbations
may be solved for analytically, as in 
\cite{Bartolo:2005kv,Matarrese:1997ay,Boubekeur:2008kn}, and seen
explicitly to be suppressed by $(\CH/k)^2$ with respect to the leading 
piece, as shown in appendix \ref{app:GRMD}, (eq. \ref{eq:GRMDsimp}):
\be
\half \delta^{(2)}_k &=&
  \left[ \left(\beta_k - \alpha_k \right)
  + { \beta_k \over 2} (\hat{q_1} \cdot \hat{q_2}) 
  \left({q_1 \over q_2}+{q_2 \over q_1}\right) 
+ \alpha_k (\hat{q_1} \cdot \hat{q_2})^2 
 + \gamma_k \left( {q_1 \over q_2}
- {q_2 \over q_2} \right)^2 \right] 
\left(  {\delta^{(1)}_{q_1} \over
1+ {3\CH^2 \over q_1^2}}
 { \delta^{(1)}_{q_2}\over 1+  {3 \CH^2 \over q_2^2} } 
 \right)\ , \nn\\
\alpha_k &=& {2 \over 7} + {59 \CH^2 \over 14 k^2} + { 45 \CH^4 \over
2 k^4 }\ , \qquad
\beta_k = 1 -  {\CH^2 \over 2 k^2 } + {54 \CH^4 \over k^4 }\ , \qquad
\gamma_k = -{3\CH^2 \over 2 k^2} + {9 \CH^4 \over 2 k^4}\ .
\ee

\begin{figure}[t!]
\begin{center}
\includegraphics[width=1.02\textwidth]{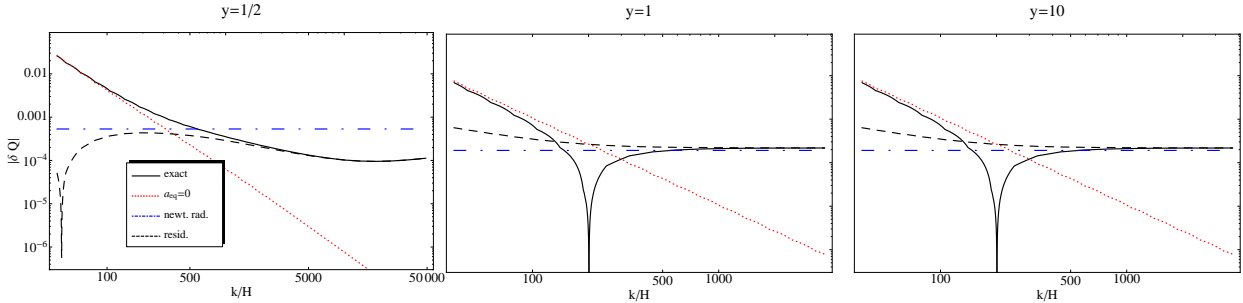}
\caption{The correction $|\delta Q|$ for $k_1=k_2=yk_3, k_3 \equiv k$ 
with $y$ fixed at
$\half, 1, 10$, from left to right, respectively.
The correction is shown for a) a matter-only universe ($a_{\rm eq} =0$),
red, dotted, b) the exact numeric correction, black, solid, and
c) the difference in Newtonian gravity between a universe with matter
and one with matter + radiation (blue, dot-dashed,
for which the log-enhanced piece of equation (\ref{eq:f2rad}) vanishes 
when $y=\half$; the size is independent of $k$ because it is the difference
between two results within Newtonian gravity).
The black dashed line is the exact radiation effect, defined as
the difference between the exact numeric correction and the matter-only
correction.
The exact numeric calculation has $a_{\rm eq} =3 \times 10^{-4}$, in which case
$k_{\rm eq}/H \approx 80$.
 }
\label{fig:deltaq2iso}
\end{center}
\end{figure}
\begin{figure}[ht!]
\begin{center}
\includegraphics[width=1.02\textwidth]{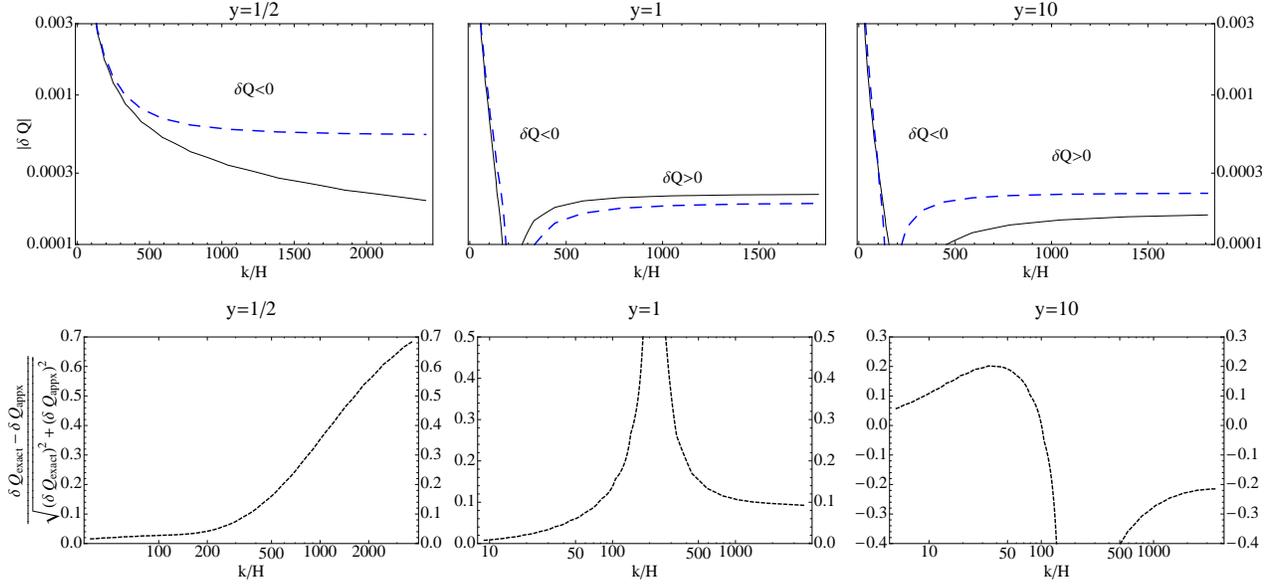}
\caption{Comparison between exact numeric corrections and an analytic
approximation $\delta Q_2 \equiv \delta Q_{2,rad} + \delta Q_{2,GR}$.
The corrections $|\delta Q|$ for $k_1=k_2=yk_3\equiv yk$ 
with $y$ fixed at
$\half, 1, 10$, for left, center, right, respectively.  In the top row, 
the exact result 
(analytic approximation) is shown in solid, black (blue, dashed). The
difference between the exact and approximate results for $\delta Q$ are
shown in the second row, divided by their sum in quadrature.  
 }
\label{fig:analyticvsexact}
\end{center}
\end{figure}
\begin{figure}[t!]
\begin{center}
\includegraphics[width=.48\textwidth]{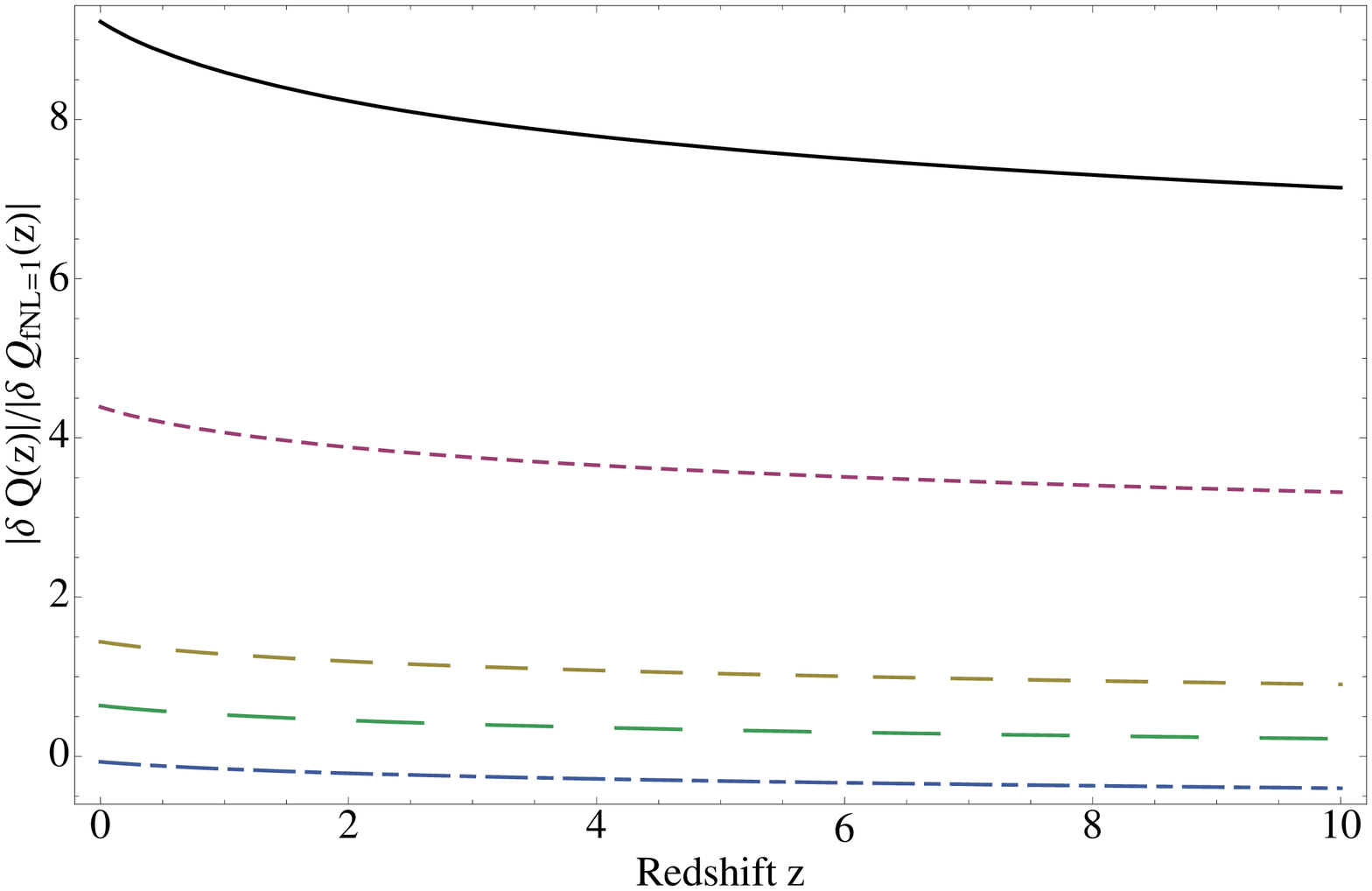}
\includegraphics[width=.48\textwidth]{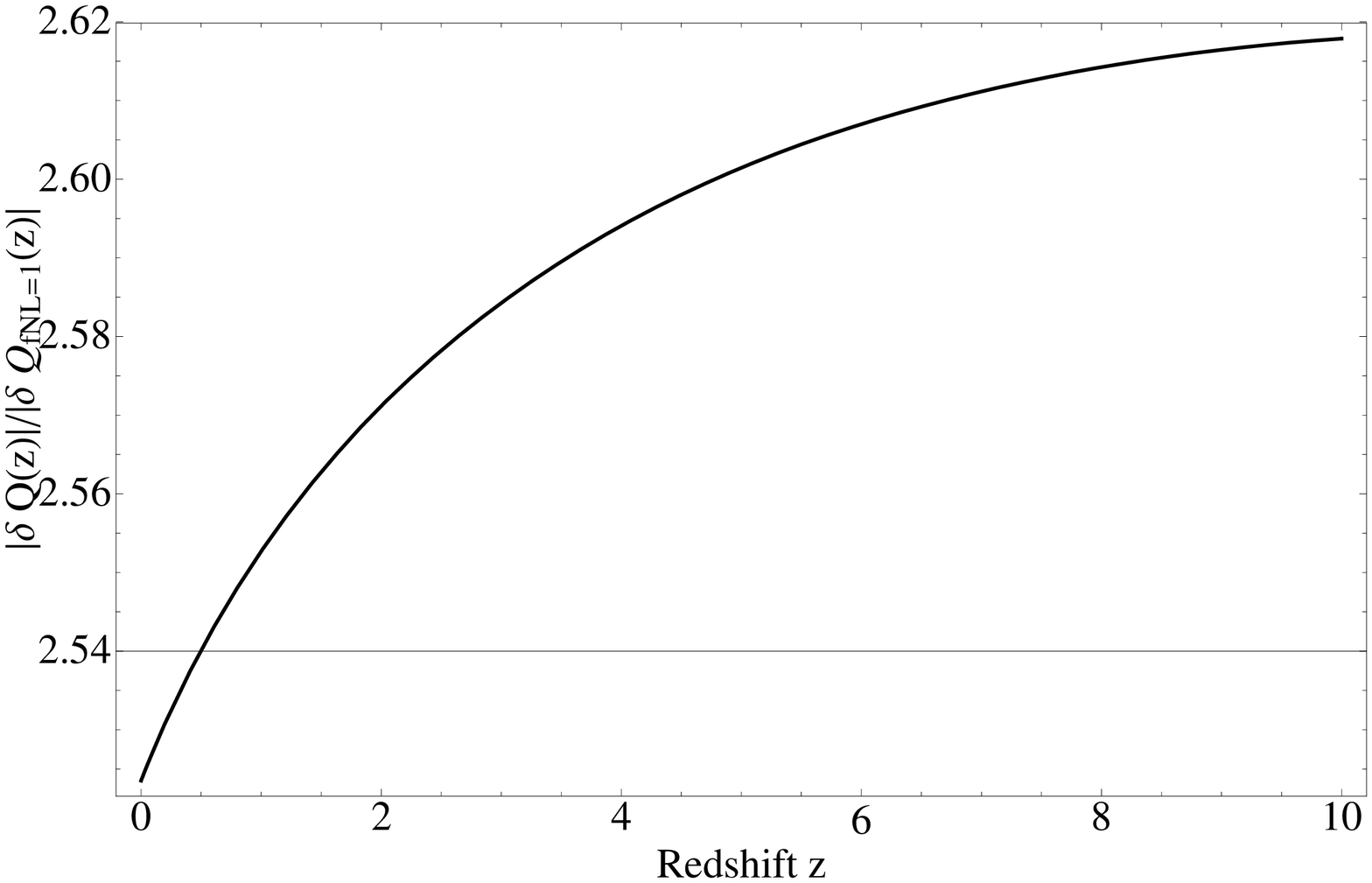}
\caption{The correction $|\delta Q|$ relative to the primordial contribution 
with $f_{\rm NL}=1$ on equilateral triangles
as a function of redshift $z$. The wavenumber in the left plot is,
from top to bottom, $k =k_{\rm eq} \times \{30, 10, 4,3, 2.3 \}$.
As $k$ decreases from $30 k_{\rm eq}$, the GR corrections grow and
approximately cancel the radiation corrections near $k\sim 2 k_{\rm eq}$.
 At smaller
$k$, the correction from GR is well-approximated by the matter-only
analytic result (\ref{eq:GRMD}).  The right plot shows the exact result
relative to $\delta Q_{f_{\rm NL} =1}$ for $k=0.4 k_{\rm eq}$, 
where essentially all the correction is from GR.  The reason the correction in
the left plot decreases at large $z$ is that $\delta/\phi$ gets smaller
in the past, so the contribution from $f_{\rm NL}$ is bigger in terms
of $\delta$.  
 }
\label{fig:vsredshift}
\end{center}
\end{figure}
 As we have discussed, however, such
effects are gauge-dependent and require more care in order
to be written in terms of present-day observables.  The effect of
radiation, on the other hand, is nearly independent of the magnitude of 
$k$ for modes that entered the horizon in the radiation era
and approaches a constant correction at small scales.  
In order to give a sense of the scale-dependence, we may fix the shape of
the triangle formed by $k,q_1$, and $q_2$. 
 In Figure \ref{fig:deltaq2iso},
we take $k_2 = k_3 = y k, k_1=k$ and plot the correction to $Q$
as a function of $k/H$.
For comparison, we also show the correction in a matter-only universe
as well as the leading correction (\ref{eq:f2rad}) 
in Newtonian gravity when radiation is included in the background (but 
radiation perturbations are neglected). For our choice of
parameters, $k_{\rm eq}/H \approx 80$, and the radiation correction
starts to dominate over PN effects between about $k_{\rm eq}$ and
$10 \ k_{\rm eq}$, depending on the triangle shape. The small
parameter suppressing the PN effects is $(H/k)^2$, and thus they
are larger than the radiation effects when $(H/k)^2 \gtrsim a_{\rm eq}
\sim (H/k_{\rm eq})^2$,
i.e. when  $k \lesssim k_{\rm eq}$.  
The radiation effects and PN effects have opposite sign for equilateral
and squeezed triangles, and thus the total correction
crosses through zero in the transition region. One can see that 
the second order perturbation $\delta^{(2)}$
approaches our simple approximation (eq. \ref{eq:f2rad}).
In fact, from Figure \ref{fig:analyticvsexact}, we see that the corrections over the entire range of $k$ are well-approximated
by the sum of the radiation corrections (eq. \ref{eq:f2rad}) and the
GR corrections (eq. \ref{eq:GRMDsimp}): $\delta Q_2 \approx 
\delta Q_{2,rad} + \delta Q_{2,GR}$.  
Note that, although the radiation correction to $Q_2$ grows 
as a function of redshift a little slower than $(1+z)$,
the correction from primordial nongaussianities has nearly the same
$z$ dependence and their ratio is nearly flat, as shown in 
Figure \ref{fig:vsredshift}.

\begin{figure}[ht!]
\begin{center}
\includegraphics[width=1.02\textwidth]{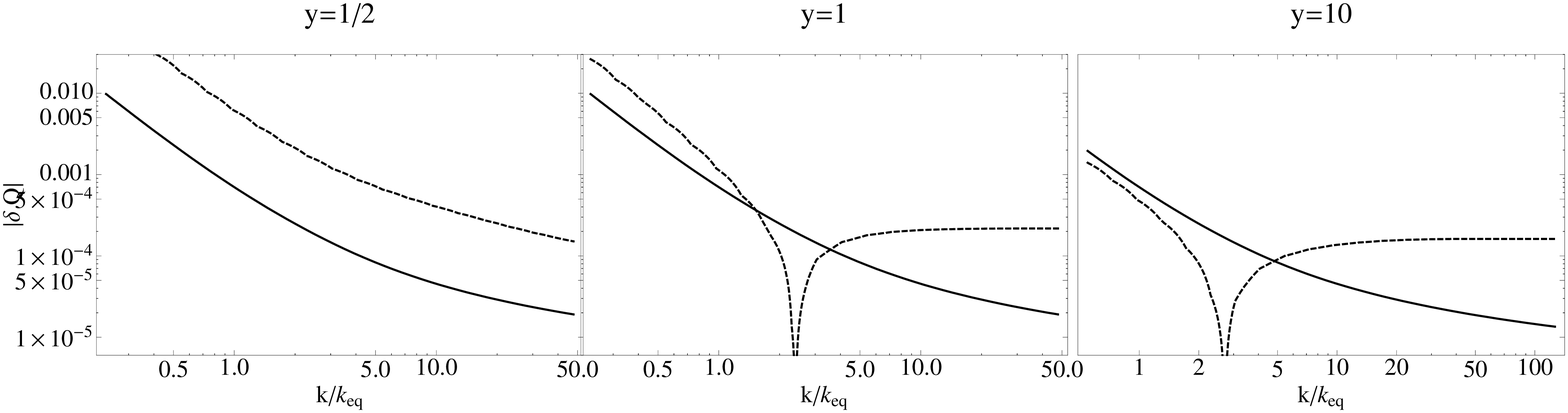}
\caption{The corrections $|\delta Q|$ for $k_1=k_2=y k_3\equiv y k$ with $y$ fixed at
$\half, 1, 10$, for left, middle right, respectively, compared to the contribution
from primordial nongaussinities with $f_{\rm NL} =1$.  
The exact numeric correction to the PT kernel is dashed, whereas
the contribution from $f_{\rm NL}$ is shown in solid.  
 }
\label{fig:vsfnl}
\end{center}
\end{figure}

Finally, we show the comparison between the corrections to $\delta Q$ and
the contribution from primordial nongaussianities with $f_{\rm NL}=1$ in
Figure \ref{fig:vsfnl}.  The correction from radiation becomes the
dominant correction only around $k\gtrsim 3 k_{\rm eq}$.
At $k= 10 k_{\rm eq}=0.13 h {\rm Mpc}^{-1}$, which is comparable to
 the largest $k$ that next-generation LSS
observations will be able to resolve \cite{Sefusatti:2007ih,Scoccimarro:2003wn},
the correction from radiation is comparable to $f_{\rm NL} \approx 4. 6$ 
for equilateral triangles.

\subsection{Including $\Lambda$}

So far, we have considered a universe with $\Omega_m + \Omega_r = 1$, 
which is only a good approximation up until $\Omega_\Lambda$ grows
to be non-negligible.  To calculate the effect of radiation
in our universe with vacuum energy 
$\Lambda \ne 0$, one simply includes the 
effect of $\Lambda$ on the background, which modifies $\CH(\eta)$.
Otherwise, the equations of motion for the linear and second order
solutions are unchanged.

\begin{figure}[ht!]
\begin{center}
\includegraphics[width=0.54\textwidth]{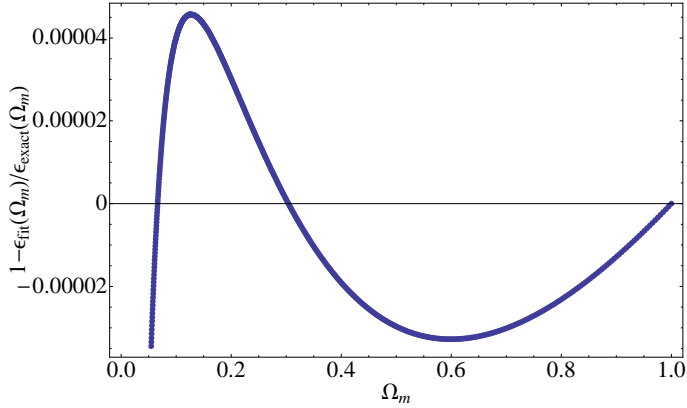}
\caption{ Comparison of the exact numeric result for $\epsilon(\Omega_m)$
compared with the parameterizaton $0.023969 (1-\Omega_m^{1/15})$ for
a flat universe with $\Lambda$ and matter only. }  
\label{fig:epsilonfit} 
\end{center}
\end{figure}

 In a universe with matter and $\Lambda$, the Newtonian equations
of motion do not admit an exact solution which is separable, i.e.
of the form $\delta^{(2)}_k(\eta) \sim F_2(q_1,q_2) 
\delta_1(q_1)\delta_1(q_2)(D^{(1)}_+(\eta))^2$.  It has been
argued that, to good approximation, they are almost separable (see e.g.
 \cite{Bernardeau:2001qr}),
and $D^{(1)}_+(\eta)$ takes the form of the (matter + $\Lambda$) linear
solutions whereas $F_2(q_1,q_2)$ is the same as the PT result. A better
approximation is to calculate $\delta^{(2)}_k$ exactly in Newtonian gravity,
matching the linear solutions onto the matter era growing modes
$\delta \propto a$ at $a \ll 1$.  When we compute $\delta^{(2)}$ numerically
in this way, we find that the result is exactly matched by the form
\be
\half F_2(q_1,q_2; \Omega_m) &=&
  \epsilon(\Omega_m) + 
{\hat{q}_1 \cdot \hat{q}_2 \over 2} \left( {q_1 \over q_2} + {q_2 \over q_1}
\right) + \left( 1- \epsilon(\Omega_m)  \right)
\left(\hat{q}_1 \cdot \hat{q}_2 \right)^2\ ,
\label{eq:f2newtlambda}
\ee
where $\epsilon(\Omega_m)$ is determined numerically.  A parameterization
of $\epsilon(\Omega_m)$ that turns out to work well is
\be
\epsilon_{\rm exact}(\Omega_m) &\approx& \epsilon_{\rm fit}(\Omega_m)
 = {5 \over 7} + c(1-\Omega_m^{1/15})\ , \qquad
c=0.023969\ .
\ee
As we show in Figure \ref{fig:epsilonfit}, this parameterization fits
the exact numeric $\epsilon(\Omega_m)$ to better than 0.0035\% for
$0.2<\Omega_m<1$.

\begin{figure}[ht!]
\begin{center}
\includegraphics[width=1.04\textwidth]{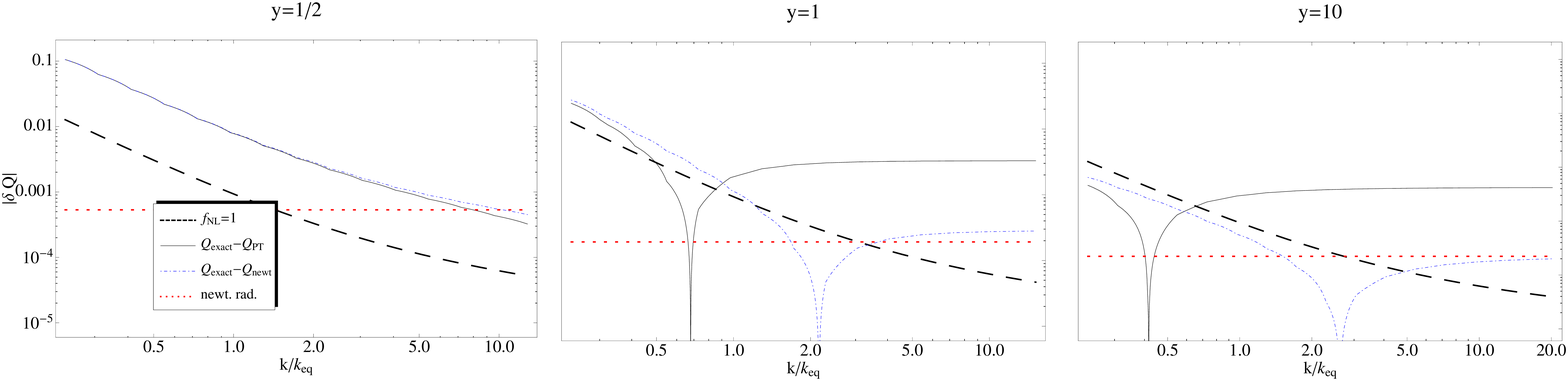}
\caption{ The corrections in a flat universe with $\Omega_{m,0}=0.27, 
\Omega_{r,0}=3 \times 10.^{-4} \Omega_{m,0}$ to $|\delta Q|$ for $k_1=k_2=y k_3\equiv y k$ with $y$ fixed at
$\half, 1, 10$, for left, middle, right, respectively, compared to the contribution
from primordial nongaussinities with $f_{\rm NL} =1$.  The black, solid line
is the difference between the exact result and the PT result.  The blue, 
dot-dashed line is the difference between the exact result and the 
Newtonian approximation of equation (\ref{eq:f2newtlambda}).
The contribution from $f_{\rm NL}=1$ is shown in dashed, black. The 
straight red, dotted line is the Newtonian radiation correction of section
\ref{sec:radnewt}, including the non-log-enhanced piece. 
The PT result is very good on squashed triangles, but off by more than
an order of magnitude more than the Newtonian approximation on equilateral
and squeezed triangles at large $k$.}
\label{fig:deltaqlambda} 
\end{center}
\end{figure}

The exact Newtonian result provides a much better approximation to the
second order density perturbations than the PT result does.  Since the
linear modes are normalized at their current values, the PT three-point
function today is insensitive to the value of $\Omega_m$.  For comparison,
in Figure \ref{fig:deltaqlambda} 
we show the difference between the PT result and the exact result for
the reduced 
three-point function, as well as the difference between the Newtonian
result and the exact result.  Including $\Lambda$, the Newtonian result
is still only off by about the equivalent of $f_{\rm NL} \approx 4$, due
to radiation.
However, the unmodified PT result of equation (\ref{eq:ptkernel}) is off by over an order of magnitude
more.

\subsection{Leading Effect of Non-Gaussianities on Galaxy Power Spectrum}

Ultimately, the dark matter three-point function must be
related to observables.  In \cite{Slosar:2008hx}, it was proposed that non-gaussianities
in the dark matter distribution might best be measured through the
additional scale-dependence they induced in the galaxy two-point
function.  This approach has the advantage that it is possible
to write the effect in terms of quantities that may be studied with
exactly gaussian statistics.  

\begin{figure}[ht!]
\begin{center}
\includegraphics[width=0.68\textwidth]{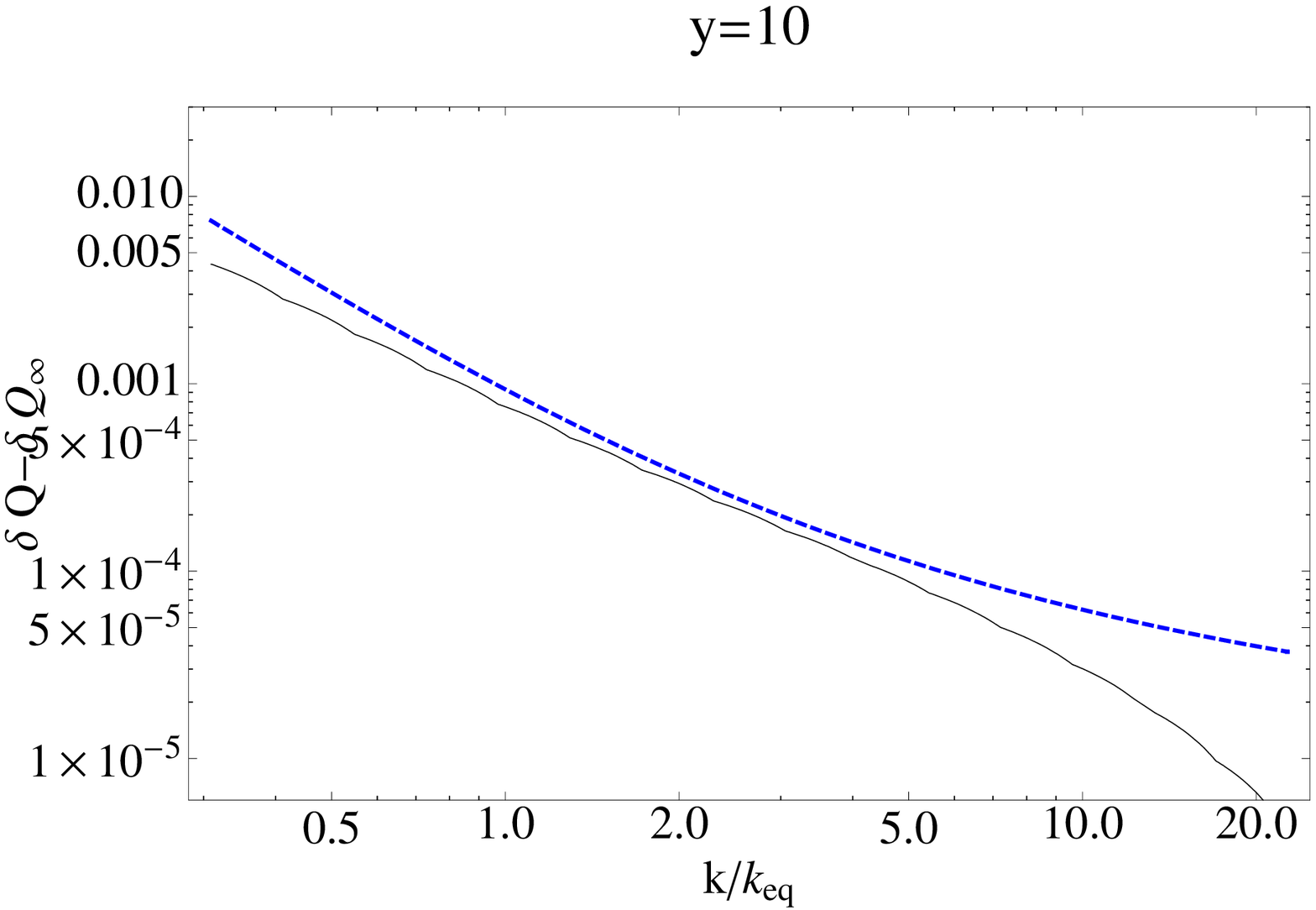}
\caption{The corrections $|\delta Q|-|\delta Q_{k\gg k_{\rm eq}}|$ 
for $k_1=k_2=10k_3$,
 compared to the contribution
from primordial nongaussinities with $f_{\rm NL} =1$.  
The correction is shown (solid, black) for the exact numeric result with $\Omega_{m,0}=0.27,
\Omega_{r,0}=3 \times 10.^{-4} \Omega_{m,0}$.
The contribution from $f_{\rm NL}$ is shown in blue, dashed. }  
\label{fig:fnlvssubleadingf2} 
\end{center}
\end{figure}

 The reason for the leading effect on the two-point function 
is that non-gaussianities correlate
the long- and short-wavelength fluctuations. With gaussian fields, a 
long-wavelength fluctuations looks like a background fluctuation that
increases or decreases the amount of short-wavelength fluctuation needed to
form a bound object \cite{Cooray:2002dia}.  However, when different modes are correlated, 
a long-wavelength fluctuation tends to increase or decrease the actual
size of the short-wavelength fluctuation. More precisely, 
with $k_L \ll k_S$,  $\delta_{k_L}$ is a long-wavelength mode that
affects the variance of the short-wavelength mode $\delta_{k_S}$:
\begin{eqnarray}
{\d  \over \d \delta_{-k_L} }\langle \delta_{k_1} \delta_{k_2} \rangle
  =
(  (\delta F_2(k_1,k_L) + {2 f_{\rm NL} \over M_{k_L}})P_{k_1}+(\delta F_2(k_2,k_L) + {2 f_{\rm NL} \over M_{k_L}})P_{k_2} ) \times \delta\left( \vec{k}_1 +
\vec{k}_2 + \vec{k}_L \right)\ , \nn \\ &&\end{eqnarray}
where the subscript $_0$ means that the expectation value is taken with $\delta_{k_L}$ equal to zero, and
where we have included the contribution from $f_{\rm NL}$ through
$\delta_{k_S} ^{(2)}= M_{k_S} ( \Phi^{(1)}_{k_S} + f_{\rm NL}(\Phi^{(1)}\star \Phi^{(1)})_{k_S})$
in linear perturbation theory, with the $\star$ indicating a convolution. We are also concentrating on the difference between the effect due to radiation and general relativity with respect to the standard newtonian treatment.   
The value of the short-wavelength mode variance determines the local value of
the rms density fluctuations, i.e. $\sigma_8$, and thus the total
effect of the long wavelength perturbation is (see for a more precise derivation \cite{Slosar:2008hx,Matarrese:2008nc}):
\be
{d n \over d \delta_{-k_L} } &=&
  \left( { d n \over d \delta_{-k_L}}\right)_0 + 
 \left( {1 \over 2P_{k_S}}
{\d  \over \d \delta_{-k_L} }\langle \delta_{k_1} \delta_{k_2} 
\rangle 
\right)  {\d n \over \d \log \sigma_8}\ .
\ee
The point of this expression is that the non-gaussianity induces a bias, and that this can be expressed in terms of the unknown quantities
$\left( { d n / d \delta_{-k_L}}\right)_0$ and 
${\d n / \d \log \sigma_8}$ that may be calculated in Newtonian gravity
with (matter+$\Lambda$) only and $f_{\rm NL} =0$. 
In the squeezed limit, 
the correction to $dn / d \delta_{k_L}$ 
from radiation approaches the value 
${8 / 35}\cdot a_{\rm eq} \log a_{\rm eq} \sin^2 \theta$ 
times $\d n / \d \log \sigma_8$ for $P(k) \propto k^{-3}$ in the 
approximation of (\ref{eq:f2rad}), where $\theta$ is the angle between
$k_L$ and $k_2$.
For isosceles triangles in the squeezed limit, $\theta = \pi/2$; the
more precise derivation in \cite{Matarrese:2008nc} demonstrates
that $\theta$ should be averaged over in the bias
with measure $\sin\theta d \theta$.
 When $k_L \gtrsim 10 k_{\rm eq}$,
the contribution to the same quantity from $f_{\rm NL}$ is
${2 f_{\rm NL} / M_{k_L}}$
times
$\d n / \d \log \sigma_8$, which is a significantly
smaller contribution 
for large $k$ and $f_{\rm NL} =1$. However, the contribution
from $f_{\rm NL}$ is $k$-dependent whereas the leading 
radiation contribution is not.  Thus, the leading radiation contribution
appears as just an overall rescaling of the galaxy bias and does not
contaminate the signal for $f_{\rm NL}$ when $f_{\rm NL}$ is fit to match
the shape of the galaxy power spectrum.  Such contamination occurs
only at sub-leading order and is comparable to $f_{\rm NL} \approx 1$. This can be seen in Figure \ref{fig:fnlvssubleadingf2}, where we show $\delta Q$, which is proportional to the bias, after subtracting off the limiting value of $\delta Q$ at large $k$, and we compare it with the analogous effect from $f_{\rm NL}=1$. Since the effect is order one, as we explained before, it becomes comparable to the GR effects that we expect to affect the distribution of collapsed objects at this order.

\section{Discussion \label{sec:discussion}}

Our main goal was to quantify the contribution from radiation
 to nongaussianities in the dark matter distribution compared with 
those from primordial nongaussianities.  We have found that the
effect on the three-point function for modes with $k \approx 10 \ k_{\rm eq}$
is comparable to $f_{\rm NL} \approx 4$.  At shorter wavelengths,
the contribution from promordial nongaussianities shrinks like
$1/\log k$ while the contribution from radiation becomes scale-independent.
Nongaussianities may also be measured through the additional scale-dependence
they induce on the power spectrum. Such scale dependence arising
from radiation corrections is comparable to $f_{\rm NL} =1$. 
We have also compared the size of primordial nongaussianities with
the relativistic corrections in Newtonian gauge to the
nonlinear growth of nongaussianities.  Such corrections are negligible
at large $k \gg k_{\rm eq}$ but at smaller than $k \lesssim k_{\rm eq}$
are comparable to $f_{\rm NL} \approx 4$ on equilateral triangles.
However, at small $k$ the effect on $\delta \rho / \rho$
 of a gauge transformation from Newtonian to synchronous gauge 
becomes large and parametrically the same as
 the relativistic corrections.  Thus, to clearly interpret this result,
$\delta \rho / \rho$ must be related to present-day observables.

We have made a number of approximations to simplify the analysis. 
We neglected baryons and neutrinos, as well as higher order
photon moments.  We also have focussed on the dark matter density
perturbations, which must eventually be related to visible objects.
Furthermore, the density field $\rho(x)$ we have been using is
the local inertial density, which will receive volume and redshift
distortions from the local metric and peculiar velocity, as well
as distortions along the line-of-sight.  If observations of
nongaussianities are to be interpreted with uncertainty
less than $\Delta f_{\rm NL} \lesssim 4$, then such distortions
most likely must be understood and quanitified as well.

\section*{Acknowledgments}
We thank S. Tassev for sharing preliminary results on second order
perturbations and D. Baumann for comments on the draft. ALF was
partially supported by DOE grant DE-FG02-01ER-40676, and an NSF graduate
research fellowship.  LS was supported in part by the National Science
Foundation  under Grant No. PHY-0503584.  MZ was supported by NASA
NNG05GJ40G and NSF AST-0506556 as well as the David and Lucile Packard,
Alfred P. Sloan and John D. and Catherine T. MacArthur foundations.

\appendix

\section*{Appendix}

\section{Second Order $\delta \rho/\rho$ in a Matter-Only Universe
\label{app:GRMD}}

The form of the second order perturbations in a matter-only universe
has been previously derived in \cite{Bartolo:2005kv,Matarrese:1997ay} using the formalism we adopt
in this paper, and also independently in \cite{Boubekeur:2008kn} using 
an action approach. The result can
be written analytically in this case.  For the convenience of the reader, we
present and comment on the results in our formalism.

In addition to setting $\Omega_r=0, \Omega_m=1$ in our equations
of motion and ignoring the radiation perturbations, 
one must also choose appropriate initial conditions. We separate out the
propagation from the initial conditions by using the method of
Green's functions, as follows.  First, it is useful
to supplement the equations of motion we have so far 
with the $G_{i0}$ component of Einstein's equation
\be
&& \CH k^2 \Psi^{(2)} + k^2 \Phi'^{(2)} - {3 \over 2} \CH^2 i k V^{(2)} =
      -S_5(k,\eta) \equiv 
\label{eq:e5}\\
&& \qquad 2 (k \cdot q_2) \Phi_{q_1}'^{(1)} \Psi_{q_2}^{(1)} 
  + 3 \CH^2 i (k \cdot \hat{q}_1)  V_{q_1}^{(1)} 
  \delta^{(1)}_{q_2} \ . \nn
\ee
For reference, we present all components of 
the Einstein tensor for a metric in Newtonian gauge 
with only scalar modes:  
\be
a^2 G^0_{\ 0} &=& -e^{-2 \Psi} 3 (\CH - \Phi')^2 
  +  e^{2 \Phi} \left( (\d \Phi)^2 - 2 \d^2 \Phi\right)\ , \\
a^2 G^i_{\ j} &=& \delta^i_j \left[ -e^{-2 \Psi} \left( (\CH - \Phi')
(\CH- 3\Phi' - 2 \Psi') + 2 (\CH' -\Phi'') \right)
  + e^{2\Phi } \left( (\d \Psi)^2 + \d^2 \Psi - \d^2 \Phi \right)
  \right] \nn\\
  && + e^{ 2\Phi} \left[ - \d_i \Psi \d_j \Psi 
  + \d_i \Phi \d_j \Phi - 2 \d_{(i} \Phi \d_{j)} \Psi  
  - \d_i \d_j (-\Phi + \Psi) \right] \ , \\
G_{i0} &=& 2 (\CH - \Phi') \d_i \Psi + 2 \d_i \Phi'\ .
\ee
It is now straightforward to obtain
a second order equation for $\delta^{(2)}$ sourced by the first order
perturbations.  We first use equation (\ref{eq:e4}) to eliminate
$\Psi^{(2)} = S_4 + \Phi^{(2)}$.  Then from (\ref{eq:e3},\ref{eq:e5}), we have
\be
V^{(2)} &=& i { 2 k^2 S_3 - 6 \CH S_5 + 3 k^2 \CH^2 \delta^{(2)} 
     + 2 k^4 \Phi^{(2)} \over 9 k \CH^3}\ .
\ee
Eliminating $\Phi'^{(2)}$ from equations (\ref{eq:e1},\ref{eq:e2}), 
we obtain $\Phi^{(2)}$ in terms of $\delta^{(2)}, \delta'^{(2)},S_i$, 
which upon substitution back into (\ref{eq:e1}) gives the desired equation
for $\delta^{(2)}$.  Using that in a matter-only universe, $\CH = 2/\eta$,
we obtain
\be
\delta^{(2)\prime \prime} + { 1296 + 72 k^2 \eta^2 - 2 k^4 \eta^4 
\over 216 \eta + 18 k^2 \eta^3 - k^4 \eta^5 } \delta^{(2)\prime}
  + { 6 k^2 (-42 + k^2 \eta^2) \over
  216 + 18 k^2 \eta^2 -k^4 \eta^4} \delta^{(2)} &=& 
   {\cal F}[S_i] \ ,\nn\\
\label{eq:delta2eq}
\ee
where ${\cal F}[S_i]$ is a function of the source terms above, and therefore
quadratic in the first order solutions. It
is fairly long~\footnote{Explicitly, 
\be
{\cal F}[S_i] &=&
  \left( k^2
   \eta  \left(k^2 \eta ^2+18\right)^2 \left(k^4 \eta ^4-18 k^2
   \eta ^2-216
\right) \right)^{-1} \times \nn\\
&& \left( -\eta ^9 S_4(\eta ) k^{12}+\eta ^9 S_3(\eta ) 
k^{10}+\eta ^9
   S_1'(\eta ) k^{10}+1188 \eta ^5 S_4(\eta ) k^8-6 \eta ^8
   S_5(\eta ) k^8\right.\nn\\
&&\left. -6 \eta ^8 S_3'(\eta ) k^8-108 \eta ^6 S_4'(\eta
   ) k^8-3 \eta ^9 S_3''(\eta ) k^8-108 i \eta ^6 S_2'(\eta )
   k^7-1188 \eta ^5 S_3(\eta ) k^6\right. \nn\\
&& \left. +3888 \eta ^3 S_4(\eta ) k^6-540
   \eta ^5 S_1'(\eta ) k^6+18 \eta ^8 S_5''(\eta ) k^6-3888 \eta
   ^3 S_3(\eta ) k^4-116640 \eta  S_4(\eta ) k^4\right. \nn\\
&& \left. +7128 \eta ^4
   S_5(\eta ) k^4-3888 \eta ^3 S_1'(\eta ) k^4+9720 \eta ^4
   S_3'(\eta ) k^4+58320 \eta ^2 S_4'(\eta ) k^4+1620 \eta ^5
   S_3''(\eta ) k^4\right. \nn\\
&& \left. -i \eta  \left(k^8 \eta ^8-1188 k^4 \eta
   ^4-3888 k^2 \eta ^2+116640\right) S_2(\eta ) k^3+58320 i \eta
   ^2 S_2'(\eta ) k^3\right. \nn\\
&& \left. -2 \left(-k^8 \eta ^8-18 k^6 \eta ^6+1620 k^4
   \eta ^4+25272 k^2 \eta ^2+69984\right) S_1(\eta ) k^2+116640
   \eta  S_3(\eta ) k^2\right. \nn\\
&& \left. +62208 \eta ^2 S_5(\eta ) k^2+93312 \eta ^2
   S_3'(\eta ) k^2+419904 S_4'(\eta ) k^2-38880 \eta ^3 S_5'(\eta
   ) k^2-+1664 \eta ^3 S_3''(\eta ) k^2\right. \nn\\
&& \left. -9720 \eta ^4 S_5''(\eta )
   k^2+419904 i S_2'(\eta ) k-279936 S_5(\eta )-419904 \eta 
   S_5'(\eta )-69984 \eta ^2 S_5''(\eta ) \right)
\ee},
but we will see that the final answer simplifies
considerably when we substitute the first order solutions.  
The above equation is
a second order sourced equation of motion for $\delta^{(2)}$ and thus
its solution is given by
\be
\delta^{(2)} &=& c_1^{(2)} \delta_1(\eta) + c_2^{(2)} \delta_2(\eta)+
   \int_0^\eta {\cal F}[S_i](\eta') G(\eta, \eta') d\eta'\ ,
\label{eq:particular}
\ee
where $G(\eta, \eta')$ is the Green's function
\be
G(\eta, \eta') &=& {\delta_1(\eta) \delta_2(\eta') - \delta_1(\eta')
\delta_2(\eta) \over
 \delta'_1(\eta') \delta_2(\eta') - \delta_1(\eta') \delta_2'(\eta')}\ ,
\label{eq:greenfunction}
\ee
and $\delta_1,\delta_2$ are the growing and decaying mode homogeneous
solutions
\be
\delta_1(\eta) &=& 12 + (k \eta)^2 \ ,\\
\delta_2(\eta) &=& {-18 + (k \eta)^2 \over (k \eta)^5}\ .
\label{eq:deltaMDsoln}
\ee
The homogeneous solutions for $\delta^{(2)}$ are the same as the
homogeneous (growing and decaying) mode solutions for $\delta^{(1)}$,
and indeed for $\delta^{(n)}$ at any order, since their equations
of motion only differ by the source term ${\cal F}$.

  At first order, the equations of
motion may easily be solved, and one finds that
$\Phi$ is constant and the matter perturbation growing mode is
\be
&& \delta^{(1)}_k = -(12+\eta^2 k^2) \left( \Phi \over 6 \right)\ , \\
&& V_k^{(1)} =  -(2 i k \eta) \left( \Phi \over 6 \right) \ .
\ee
Upon substituting these into the source term ${\cal F}$, the Green's
function integral may be performed analytically:
\be
\half \int_0^\eta {\cal F}[S_i](\eta') G(\eta,\eta') d\eta' &=& \left[
\left(\frac{k^4}{14}+\frac{3}{28} \left(q_1^2+q_2^2\right)
   k^2-\frac{5}{28} \left(q_1^2-q_2^2\right)^2\right) \eta
   ^4 \right. \nn\\
&& \left. +\left(-\frac{23 k^2}{7}-\frac{167}{14}
   \left(q_1^2+q_2^2\right)+\frac{45
   \left(q_1^2-q_2^2\right)^2}{14 k^2}\right) \eta ^2 \right]
\left( {\Phi^{(1)}_p(q_1) \Phi^{(1)}_p(q_2) \over 36} \right) \nn\\
\ee
where we have factored out $\Phi^2/36$ in order to more easily compare
the leading term with the PT kernel.
To obtain the correct initial conditions, we take the early-time
super-horizon limit $\eta\rightarrow 0, k\rightarrow 0$ as 
before (see paragraph preceding equation (\ref{eq:secondorderinit})):
\be
&&\delta_m^{(2)} 
 = - {2 \over 3 \CH^2} S_3 - 2 S_4- 2 \Phi^{(2)} \ ,\nn\\
&& \Psi^{(2)} = \Phi^{(2)} + S_4\ .
\label{eq:secondorderinitmatter}
\ee
Outside the horizon during the matter era, $\zeta = -\Phi - {2 \over 3}
\Psi = -{5 \over 3} \Phi - {1 \over 3} S_4$, and
thus  $\Phi^{(2)} = -{2 \over 5} S_4$ initially.  
Using the limiting values of $S_3=-6 \CH^2 \Phi^{(1)}_{q_1}
\Phi^{(1)}_{q_2}, S_4 = -{10 \over k^2}((\hat{k} \cdot q_1 \hat{k} \cdot q_2
-{1 \over 3} q_1 \cdot q_2) \Phi_{q_1}^{(1)} \Phi_{q_2}^{(1)}$, we obtain
the initial condition for $\delta^{(2)}$:
\be
\delta^{(2)}_{\rm init} &=&
  \left( { 5 k^4 -3(q_1^2 -q_2^2)^2 +2 k^2(q_1^2 + q_2^2) \over k^4} \right)
\Phi^{(1)}_{q_1} \Phi^{(1)}_{q_2}\  .
\ee
Since the initial conditions are set at $\eta \rightarrow 0$,
there is no component of decaying mode in equation (\ref{eq:particular}),
and thus $c_1 = \delta^{(2)}_{\rm init}/12$.  Putting everything together,
we finally obtain
\be
\half \delta^{(2)}_k &=& 
 \left[
\left(\frac{k^4}{14}+\frac{3}{28} \left(q_1^2+q_2^2\right)
   k^2-\frac{5}{28} \left(q_1^2-q_2^2\right)^2\right) \eta
   ^4 \right. \nn\\
&& \left. +\left(\frac{59 k^2}{14}-\frac{125}{14}
   \left(q_1^2+q_2^2\right)-\frac{9
   \left(q_1^2-q_2^2\right)^2}{7 k^2}\right) \eta ^2  \right. \nn\\
&& \left. + \left(90 + 36 {q_1^2 + q_2^2 \over k^2}
-54 {(q_1^2 -q_2^2)^2 \over k^4} \right) \right]
\left( {\Phi^{(1)}_p(q_1) \Phi^{(1)}_p(q_2) \over 36} \right) \ .
\label{eq:GRMD}
\ee
This agrees with the expression for $\delta \rho/\rho$ one may obtain
from the second-order metric in \cite{Boubekeur:2008kn}.  We may rewrite this
in a slightly different form to more easily compare with the
PT result of equation (\ref{eq:ptkernel}), which corresponds to the
$k\eta \rightarrow \infty$ limit:
\be
&&\half \delta^{(2)}_k =
  \left[ \left(\beta_k - \alpha_k \right)
  + { \beta_k \over 2} (\hat{q_1} \cdot \hat{q_2}) 
  \left({q_1 \over q_2}+{q_2 \over q_1}\right) 
+ \alpha_k (\hat{q_1} \cdot \hat{q_2})^2 
 + \gamma_k \left( {q_1 \over q_2}
- {q_2 \over q_2} \right)^2 \right] \nn\\
&& \qquad \qquad \qquad 
\times \left(  {\delta^{(1)}_{q_1} \over
1+ {3\CH^2 \over q_1^2}}
 { \delta^{(1)}_{q_2}\over 1+  {3 \CH^2 \over q_2^2} } 
 \right)\ , \nn\\
&&\alpha_k = {2 \over 7} + {59 \CH^2 \over 14 k^2} + { 45 \CH^4 \over
2 k^4 }\ , \qquad
\beta_k = 1 -  {\CH^2 \over 2 k^2 } + {54 \CH^4 \over k^4 }\ , \qquad
\gamma_k = -{3\CH^2 \over 2 k^2} + {9 \CH^4 \over 2 k^4}\ .
\label{eq:GRMDsimp}
\ee

\subsection{Squeezed Limit of $\delta^{(2)}$}

We note one immediate check of equation (\ref{eq:GRMD}) based on the
``squeezed'' limit $q_1 \sim k \gg q_2$, at leading order.  In this limit, $\Phi_{q_2}$
is a very long wavelength mode whose only physical effect
is to modify the physical time and wavelength of the short wavelength
mode \cite{Maldacena:2002vr,Creminelli:2004yq,sexybeasts}.  More precisely, coordinates $(\eta, x)$ with a long wavelength 
background 
$\Phi_{k_l}$ are equivalent to the modified coordinates $(\eta',x')$, where
\be
&&a(\eta') dx' = (1-\Phi_{k_l}) a(\eta)dx\ , \\
&& a(\eta') d\eta' = (1+ \Phi_{k_l}) a(\eta) d\eta\ .
\ee
During the matter-era, $a(\eta) \propto \eta^2$, and thus 
\be
&&\eta' = \eta (1+ {1 \over 3} \Phi_{k_l}) \ , \\
&& x' = x (1-{5 \over 3} \Phi_{k_l})\ .
\ee
The effect of this coordinate redefinition on the short
wavelength matter density $\rho_{k_s}$
is 
\be
\rho_{k_s} &\rightarrow& \rho_{k_s} + {1 \over 3} \Phi_{k_l}
  {\d \rho_{k_s} \over \d \log \eta}
+ {5 \over 3} \Phi_{k_l} {\d \rho_{k_s} \over \d \log {k_s}} \nn\\
 &=& \bar{\rho} \delta_{k_s} + \bar{\rho} 
{1 \over 3} \Phi_{k_l} \left( -3 \eta \CH \delta_{k_s} + \eta \delta'_{k_s} \right)
+ {5 \over 3 } \Phi_{k_l} \bar{\rho} k_s {\d \delta_{k_s} \over \d k_s} \nn\\
&=& \bar{\rho} \left( \delta + 2 (72 -6 k_s^2 \eta^2) 
\left( \Phi_{k_l} \Phi_{k_s} \over 36 \right) \right)\ .
\label{eq:squeezedargument}
\ee

In terms of the explicit expression for $\delta^{(2)}$ (eq. \ref{eq:GRMD}), the effect
of the long wavelength mode $\Phi_{k_l}$ is obtained from 
$\half \delta^{(2)} \rightarrow \half \delta^{(2)}_{k_s} (q_1 =k_s, q_2 = k_l) + \half \delta^{(2)}_{k_s}
(q_1=k_l, q_2=k_s)$ in the limit $k_l \ll k_s$, and we find
\be 
&&\half \delta^{(2)}_{k_s}  \rightarrow 
2(72 -6 k_s^2 \eta^2) \left( \Phi_{k_l} \Phi_{k_s} \over 36 \right)\ ,
\ee
which indeed agrees with equation (\ref{eq:squeezedargument}).

\end{document}